\documentclass[english,12pt,a4paper]{article}

\usepackage{authblk}
\usepackage[text={17.0cm,23.7cm}]{geometry}
\usepackage[T1]{fontenc}
\usepackage[utf8]{inputenc}
\usepackage{babel}
\usepackage{hyperref}
\usepackage{cite}
\usepackage{graphicx}
\usepackage{amsmath}
\usepackage{amssymb}
\usepackage{xcolor}
\usepackage{ulem}
\usepackage[font=small,labelfont=bf]{caption}

\pdfoutput=1

\title{Current status of the project}

\newcommand{\snu}{\tilde{\nu}}

\newcount\mstylenum 
\def\varstyle#1{\mathchoice{\mstylenum=0 #1}{\mstylenum=1 #1}{\mstylenum=2 #1}{\mstylenum=3 #1}}
\def\usestyle{\ifcase\mstylenum \displaystyle\or\textstyle\or\scriptstyle\or\scriptscriptstyle\fi}

\def\gensim#1{\mathrel{\varstyle{\vtop{\offinterlineskip
   \halign{\hfil$\usestyle##$\hfil\cr#1\cr\noalign{\kern1pt}\sim\cr}}}}}
\def\lessim{\gensim<}

\usepackage{color}
\definecolor{myred}{rgb}{0.6,0,0} %usage:  {\textcolor{myred}{Hello World}}
\definecolor{myblue}{rgb}{0,0.2,0.4}
\definecolor{mygreen}{rgb}{0,0.9,0.1}
\definecolor{hc}{rgb}{.9,0.1,0.7}
\definecolor{hcout}{rgb}{.9,0.7,0.9}
\definecolor{Orange}{rgb}{1.,0.65,0.}
\begin{document}
\title{\vspace{-2cm}
{\normalsize
\flushright HRI-RECAPP-2019-007, TUM-HEP 1224/19,\,KIAS-P19056\\}
\vspace{0.6cm}
\textbf{Gamma-ray signals from multicomponent scalar dark matter decays}\\[8mm]}

\author[1]{Avirup Ghosh}
\author[2,3]{Alejandro Ibarra}
\author[1,3]{Tanmoy Mondal}
\author[1]{Biswarup Mukhopadhyaya 
\thanks{ Address after December 2: Department of Physical Sciences, Indian Institute of Science Education and Research, Kolkata, Mohanpur - 741246, India.}}
\affil[1]{\normalsize\textit{Regional Centre for Accelerator-based Particle Physics,Harish-Chandra Research Institute, HBNI, Chhatnag Road, Jhunsi, Allahabad - 211 019, India}}
\affil[2]{\normalsize\textit{Physik-Department, Technische Universit\"at M\"unchen, James-Franck-Stra\ss{}e, 85748 Garching, Germany}}
\affil[3]{\normalsize\textit{School of Physics, Korea Institute for Advanced Study, Seoul 02455, South Korea}}

\date{}

\maketitle
\begin{abstract}

Within a multicomponent dark matter scenario, novel gamma-ray signals may 
arise from the decay of the heavier dark matter component into the 
lighter. For a scalar dark sector of this kind, the decay $
\phi_2\rightarrow\phi_1 \gamma$ is forbidden by the conservation of 
angular momentum, but the decay $\phi_2 \rightarrow \phi_1 \gamma\gamma$ 
can have a sizable or even dominant branching ratio. In this paper we 
present a detailed study of this decay channel. We determine the width 
and photon energy spectrum generated in the decay, employing an effective 
theory approach, and in UV complete models where the scalar dark matter 
components interact with heavy or light fermions. We also calculate 
limits on the inverse width from current data of the isotropic diffuse 
photon flux, both for a hierarchical and a degenerate dark matter spectrum. 
Finally, we briefly comment on the prospects of observing the diphoton 
signal from sneutrino decay in the minimal supersymmetric standard model 
extended with right-handed neutrino superfields ($\snu$MSSM). 
\end{abstract}

\section{Introduction}

There is mounting evidence that approximately 80\% of the matter content 
of the Universe is not in the form of baryons, electrons or neutrinos~\cite{Ade:2015xua}.
An exciting hypothesis is that this mysterious form 
of matter, usually dubbed dark matter, is constituted by new particles 
not contained in the Standard Model (for reviews, see {\it e.g.} \cite{Bertone:2004pz,Bergstrom:2000pn,Feng:2010gw}).
However, up to this day, this hypothesis remains unproved. 

A possible strategy to establish the particle nature of the dark matter 
consists in the search for the products of dark matter annihilation or 
decay, either in the form of photons, antimatter particles or neutrinos. 
This search is challenging, due to the existence of large (and not always 
sufficiently well understood) astrophysical backgrounds. On the other 
hand, in a given dark matter framework the intensity and energy spectrum 
of the products of annihilation or decay can be calculated, thus 
permitting in principle a dedicated search for this exotic component in 
the data. In many cases, the exotic flux component is expected to have a 
smooth energy spectrum, which is not easily distinguishable from the 
background. However, some dark matter frameworks predict fluxes at Earth with energy spectra which are
distinctively different from the ones predicted by state-of-the-art 
background models. For these frameworks, current instruments can be very 
sensitive to annihilation or decay signals. More importantly, the 
identification of such distinctive signal would hence constitute an 
evidence for the particle nature of the dark matter. 

Generically, these distinctive features appear in dark matter 
annihilations or decays where the final state contains stable particles 
with energies at, or close to, the kinematic cut-off. More concretely, 
distinctive features in the photon spectrum arise in the two-body final 
state $\gamma {\cal N}$ (with ${\cal N}=\gamma, Z, h, \nu$ a neutral 
Standard Model particle), which produces a gamma-ray line~\cite{Pal:1981rm,Srednicki:1985sf,Bergstrom:1988fp}, 
and in the three-body final states $\bar f f \gamma$~\cite{Bergstrom:1989jr,Flores:1989ru,Bringmann:2007nk} (with $f$ a charged fermion) 
and $W^+W^-\gamma$~\cite{Garcia-Cely:2013zga} provided the effective interaction inducing the process 
includes the photon. The large exposure and excellent energy resolution 
of the AMS-02 electron and positron data allow to search for the spectral 
feature produced in the two-body final state $e^-e^+$~\cite{Bergstrom:2013jra,Ibarra:2013zia} and in the three-body final state $e^- e^+ {\cal 
N}$~\cite{Bergstrom:2013jra}, and to a lesser extent for the same 
processes with muons or taus in the final state. Finally, the fairly good 
energy resolution for the cascade events at IceCube opens up the possibility 
of observing features in the neutrino energy spectrum from the two body 
decays $\nu {\cal N}$~\cite{Aisati:2015vma}.

It is plausible that the dark sector contains more than one particle. If this is the case, novel and distinctive features could be detected. 
For example, if the final state contains two new particles that decay in flight into photons, the resulting photon energy spectrum presents 
a box-like shape~\cite{Ibarra:2012dw}, a triangle-like shape~\cite{Ibarra:2016fco}, or in general a polynomial shape~\cite{Garcia-Cely:2016pse}. 

Furthermore, there could be more than one particle contributing to the dark matter density. 
In this class of scenarios, there may be additional dark matter signals from the decay of a heavier dark matter component into a lighter, 
if allowed by the symmetries of the model. Of special interest is the case where the mass difference is small, possibly due to the mild breaking of a dark sector symmetry, 
such that the phase space available in the decay is small, thus leading to longer lifetimes (in analogy to the slow neutron decay into a proton, electron and antineutrino, 
which is a consequence of the mild breaking of the isospin symmetry). For sufficiently small mass differences, only decays into photon and neutrinos would be kinematically accessible, 
thus naturally leading to distinctive signals in the cosmic fluxes of these particle species. 

More concretely, for multicomponent fermion dark matter the decay $\psi_2\rightarrow \psi_1\gamma$  
would produce a line in the photon spectrum at an energy $E_\gamma=M_2/2(1-M_1^2/M_2^2)$, with $M_1$ and $M_2$ the 
masses of the dark matter components $\psi_1$ and $\psi_2$. The signals in this case, are analogous to those from a 
single-component fermionic dark matter scenario with decay $\psi\rightarrow \gamma\nu$, and which generates a photon 
with energy $E_\gamma=m_\psi/2$, and which has been thoroughly studied in the literature. This class of models generically 
predicts also the three-body decays $\psi_2\rightarrow \psi_1 \nu\bar\nu$ and, if kinematically allowed,  $\psi_2\rightarrow \psi_1 e^+e^-$ or 
into other charged fermions, which contribute respectively to the neutrino flux or to the electron/positron flux.

The case of the multicomponent scalar dark matter has received less attention (see, however, \cite{Essig:2013goa}). 
The two body decay $\phi_2\rightarrow \phi_1\gamma$ is forbidden by the conservation of total angular momentum. 
On the other hand, the process  $\phi_2\rightarrow \phi_1\gamma\gamma$ is allowed. In this paper we will study this process in detail, 
focusing on the case where the mass difference between the two dark matter components is small. In this regime, the photon spectrum produced in the decay has a 
distinctive shape that allows a sensitive search for this signal in the gamma-ray data. Furthermore, the branching ratio of this process can be sizable (or even dominant). 
Hence, the search for the photon signals  would constitute the most powerful probe of this scenario.

The paper is organized as follows: In section~\ref{sec:eff_theory} we consider an effective theory approach to describe the decays
$\phi_{2}\rightarrow\phi_{1}\gamma\gamma$, $\phi_{2}\rightarrow\phi_{1}f\bar{f}$, 
and we provide a simple model where the effective interactions are generated at the one loop level due to a Yukawa coupling of the 
dark matter components with two exotic electrically charged heavy fermions. In  Section~\ref{sec:coupling-to-SM} we consider a variant of this model, 
where the two dark matter components interact with one light Standard Model fermion and one heavy exotic fermion, in which case the effective theory 
description of the decay $\phi_2\rightarrow \phi_1\gamma\gamma$ may not be valid. In section~\ref{sec:right_snu_DM} we consider a  concrete realization 
of the latter scenario in the context of the $\snu$MSSM. Finally, in section~\ref{sec:conclusion} we summarize and conclude.

\section{Effective theory approach to multicomponent scalar dark matter decay}
\label{sec:eff_theory}

We consider a scenario where the Standard Model is extended with two scalar gauge singlets, $\phi_1$ and $\phi_2$, with masses $M_1$ and $M_2$ respectively. We also introduce a $Z_2$ symmetry, under which $\phi_1$ and $\phi_2$ are odd, while all Standard Model particles are even. Therefore, interaction terms of the form $\phi_i |H|^2$ are forbidden. We assume that $\phi_1$ is the lightest particle of the $Z_2$-odd sector. 
Then, $\phi_1$ is cosmologically stable and constitutes a dark matter candidate. The heavier scalar singlet $\phi_2$ decays into $\phi_1$, but it is assumed to be long-lived in cosmological time-scales. 
In this scenario, therefore, the dark matter contains two components with abundances $\Omega_{\phi_1} h^2$ and $\Omega_{\phi_2} h^2$. 
The relic density of both dark matter components can be determined {\it e.g.}  by thermal freeze-out \cite{Silveira:1985rk,McDonald:1993ex} or by thermal freeze-in \cite{Yaguna:2011qn}
depending on the model parameters. In what follows, we will not address dark matter production, but simply assume that the total dark matter abundance is 
$(\Omega_{\phi_1}+\Omega_{\phi_2}) h^2\simeq 0.12$, as determined by the Planck satellite~\cite{Ade:2015xua}. 

The signals of the mono-component singlet scalar dark matter scenario have been thoroughly discussed in the literature, 
and the extension to the multi-component variant of the model is straightforward. 
In this work, therefore, we will focus on the aspects of the model that are specific to the multicomponent character of our framework. 
Concretely, we will focus on the signals arising from the decay of the heavier $Z_2$-odd dark matter component into the lighter. 
The decay can be induced by the Higgs portal term 
\begin{align}
-{\cal L}_{{\rm dim}-4}=f_3 \phi_2 \phi_1 |H|^2\;,
\label{eq:Higgs_portal}
\end{align}
or by dimension six operators of the form
\begin{align}
-{\cal L}_{{\rm dim}-6}=\frac{g_4}{\Lambda_4^2} \phi_2 \phi_1 {\cal S}_{{\rm dim}-4}+\frac{g_3}{\Lambda_3^2}  \phi_2 \partial_\mu \phi_1  {\cal V}_{{\rm dim}-3}^\mu+\frac{g'_3}{\Lambda_3^2}\phi_1 \partial_\mu \phi_2  {\cal V}_{{\rm dim}-3}^\mu+\frac{g_2}{\Lambda_2^2} \partial_\mu \phi_1 \partial_\nu \phi_2 {\cal T}^{\mu\nu}_{{\rm dim}-2}\;,
\label{eq:gen_eff_op}
\end{align}
where ${\cal S}_{{\rm dim}-4}$, ${\cal V}^\mu_{{\rm dim}-3}$ and ${\cal T}^{\mu\nu}_{{\rm dim}-2}$ are, respectively, any gauge invariant dimension-four scalar, dimension-three vector or dimension-two tensor operator involving Standard Model particles only. Besides, $\Lambda_i$ denotes the typical mass scale of the particles generating the corresponding effective interaction, and $g_i$ are dimensionless parameters; the validity of our effective theory requires $\Lambda_i\gg M_1,M_2$.

\begin{figure}[t!]
	\begin{center}
		\includegraphics[width=0.8\textwidth]{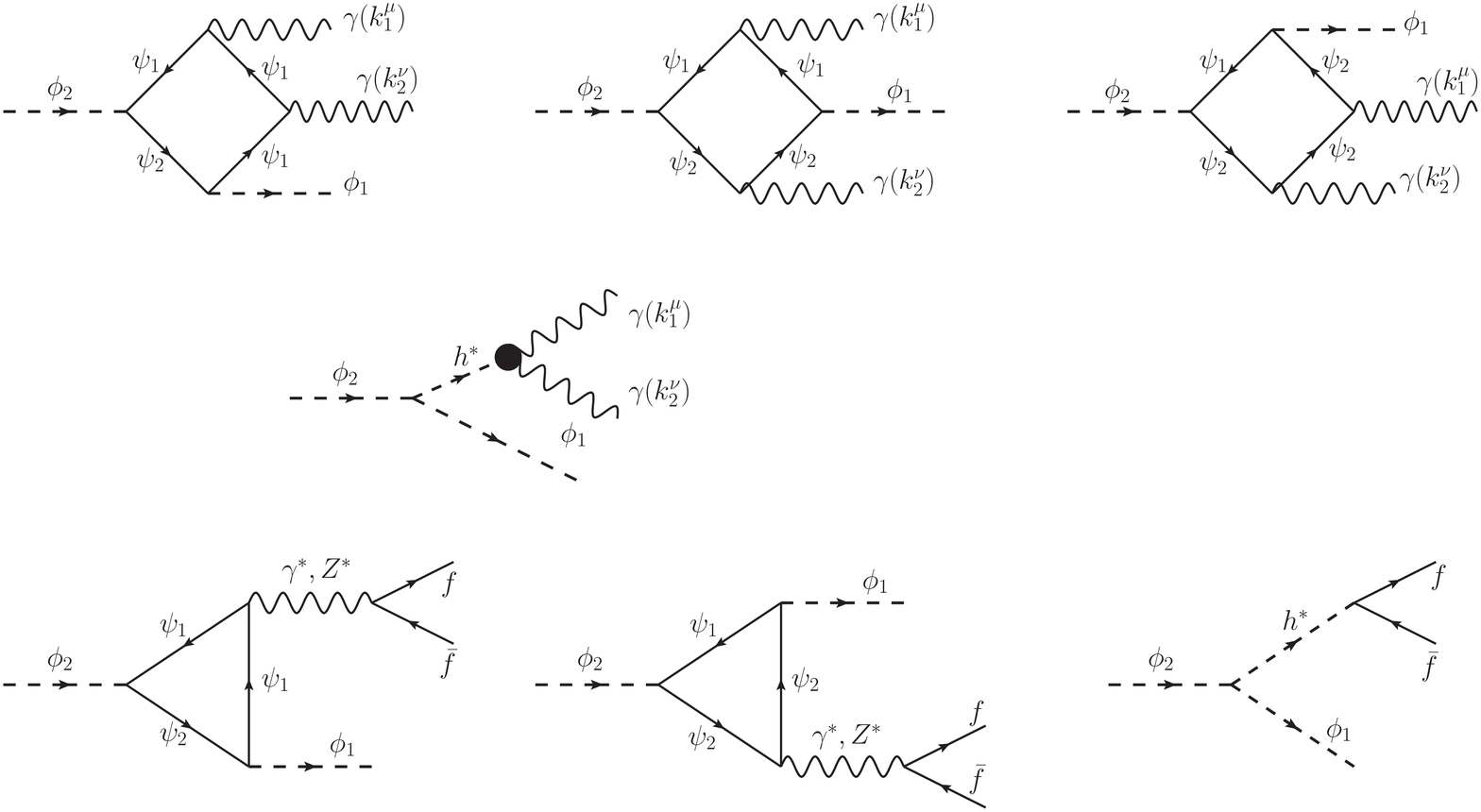}
		\caption{One-loop Feynman diagrams contributing to the processes 
			$\phi_{2}\rightarrow\phi_{1}\gamma\gamma$  and $\phi_{2}\rightarrow  \phi_{1}f\bar{f}$. Diagrams where 
			the photon lines are interchanged (not shown in the Figure) also contribute to the amplitudes.}
		\label{fig:feyndiags}
	\end{center}	
\end{figure}

These effective interactions could be generated, for instance, by extending the model with heavy vector-like fermionic fields $\psi_1$($Z_2$-even) 
and $\psi_2$($Z_2$-odd) with masses $m_1$ and $m_2$, respectively, singlets under $SU(3)_c\times SU(2)_L$ and with hypercharge $-1$, which couple to the scalar field $\phi_i$ 
via a Yukawa interaction $Y_i \phi_i \overline \psi_1 \psi_2$. 
Integrating out the heavy fermions, one obtains the following dimension-six operators involving the electromagnetic field strength tensor, through the diagrams shown in  Fig.~\ref{fig:feyndiags}:
\begin{align}
-\mathcal{L}_{\rm int}\,\supset&\,\frac{f_{1}}{\Lambda^{2}}\,\left(\partial_{\mu}\phi_{2}\partial_{\nu}\phi_{1}-\partial_{\nu}\phi_{2}\partial_{\mu}\phi_{1}\right)\,F^{\mu\,\nu}\,+\,\frac{f_{2}}{\Lambda^{2}}\phi_{2}\phi_{1}\,F^{\mu\,\nu}F_{\mu\,\nu}\;, \label{eq:eff-Lag}
\end{align}
with
\begin{align}
\frac{f_1}{\Lambda^2}&\simeq \frac{\alpha_{\rm EM}^{1/2} }{12 \pi^{3/2}m_1 m_2} g\left(\frac{m_1}{m_2}\right) {\rm Im}(Y_1 Y_2^*),\nonumber\\
\frac{f_2}{\Lambda^2}&\simeq \frac{ \alpha_{\rm EM}}{12\pi\,m_{1}\,m_{2}} {\rm Re}(Y_1 Y_2^*).
\label{eqn:op_toymodel_m1nem2}
\end{align}
Here,
\begin{align}
g(x)=\frac{3x\left(1-4x+x^2\right)}{\left(1-x^2\right)\,\left(1-x\right)^2}+\frac{4x\left(1-3x+x^2-3x^3+x^4\right)\log x}{\left(1-x^2\right)^2\,\left(1-x\right)^2}
\end{align}
is a function that satisfies $g(x)=-g(x^{-1})$ and which vanishes at $x=1$ ({\it i.e.} when $m_1=m_2$) and when $x\gg 1$ or $x\ll 1$ ({\it i.e.} when there is a large hierarchy between $m_1$ and $m_2$); the vanishing of the Wilson coefficient $f_1=0$ when $m_1=m_2$ is due to Furry's theorem, as in this limit the vertex factors remain invariant under the reversal of the fermion directions in the loop.
 Note also that $f_1$ is non-zero only when the relative phase between $Y_1$ and $Y_2$ is different from $0$ or $\pi$. Analogous interactions involving the $Z$-boson arise upon replacing $A^\mu\rightarrow -\tan\theta_{\rm W} Z^\mu$, with $\theta_{\rm W}$ being the Weinberg's angle.

The possible two body decay final states include a Standard Model neutral boson, either a photon, a $Z$ boson or a Higgs boson. 
It can be checked that the decay rate for $\phi_2\rightarrow\phi_1 \gamma$ via the operator proportional to $f_1$ is zero for an on-shell photon, 
in concordance with the conservation of angular momentum. On the other hand, decays involving one $Z$-boson or one Higgs boson are allowed. 
The signatures of these decays are identical to those produced by the well studied decays $\phi\rightarrow ZZ, hh$, with the appropriate shift in the energy of the $Z$ or the Higgs boson. 

 In this work we concentrate in a scenario where the two-body decays $\phi_2\rightarrow \phi_1 X$,  with $X$ a massive boson, are kinematically forbidden. One can define the parameter $\Delta\equiv 1-M_1^2/M_2^2$, which measures the mass degeneracy between the two dark matter components: $\Delta\simeq 1$ corresponds to a very hierarchical spectrum and $\Delta\simeq 0$ to a very degenerate spectrum. The condition $M_2< M_1 + M_X$ that forbids the decay $\phi_2\rightarrow \phi_1 X$, with $M_X$ the mass of the boson $X$, then translates into $M_2\leq M_X/ (1-\sqrt{1-\Delta})$. This condition is satisfied, in particular for the hierarchical spectrum when $\phi_2$ is lighter than $X$, and in the degenerate limit when $M_2\lesssim 2M_X/\Delta$.~\footnote{For small mass splittings, the effective operators of Eq.~\eqref{eq:gen_eff_op} 
in general should contain not quarks and gluons as degrees of freedom but pions, as dictated by chiral perturbation theory. 
In this regime, the two body decay $\phi_2\rightarrow\phi_1 \pi^0$ or similar decays involving mesons may occur.} 
%{\color{red}\sout{Therefore, the most interesting regions for these three-body final states containing diphotons is $M_2-M_1< m_{\pi^0}$.} *** Better not to say this; the pion decays into two photons***}} 
The possible three body decays are  then $\phi_2\rightarrow \phi_1 f\bar f$, with $f$ being a Standard Model fermion, and $\phi_2\rightarrow \phi_1 \gamma\gamma$. 
While three body decays with two fermions in the final state have been discussed in the literature in other scenarios, the three body decay with two photons 
in the final state has received less attention (see, however, \cite{Essig:2013goa}).

The decay $\phi_2\rightarrow\phi_1 f\bar f$, if kinematically allowed, is induced by the mediation of an off-shell photon 
(via the dimension-six operator proportional to $f_1$ in Eq.~(\ref{eq:eff-Lag})), an off-shell $Z$-boson (via the dimension-six 
operator proportional to $f_1$ in Eq.~(\ref{eq:eff-Lag}), replacing $A^\mu\rightarrow -\tan\theta_{\rm W}Z^\mu$), 
and an off-shell Higgs boson (via the dimension-four operator proportional to $f_3$ in Eq.~(\ref{eq:Higgs_portal})). The differential decay rate reads
\begin{align}
\frac{d\Gamma_{\phi_{1}f\bar{f}}}{dx_f}=&\,\frac{f^{2}_{1}N_c \alpha_{\rm EM}}{192\pi^3\Lambda^{4}} M_2^5\Delta^5\frac{x_f^2(1-x_f)^{2}}{(1-x_f\Delta)}\left[12 q_f^2+8q_f c_v^{(f)}\frac{M_2^2 \Delta^2}{\cos^2\theta_{\rm W} m^{2}_{Z}}\frac{x_f(1-x_f)}{(1-x_f\Delta)}\right.
\nonumber\\ &\left.+(c_v^{(f)2}+c_a^{(f)2})\frac{M_2^4 \Delta^4}{\cos^4\theta_{\rm W} m^{4}_{Z}}\frac{x_f ^2(1-x_f)^2}{(1-x_f\Delta)^2}\right]
\,+\,\frac{f^{2}_{3}N_c m^{2}_{f}}{128\pi^{3}m^{4}_h}M^{3}_{2}\Delta^{5}\frac{x_f^2(1-x_f)^{2}}{(1-\Delta\,x_f)^{2}},
\label{eqn:DGammaffbar}
\end{align}
where we have defined $x_f\equiv\frac{2E_{f}}{M_{2}\Delta}$, 
which is kinematically restricted to be in the range $0\leq x_f \leq 1$. Here, $m_f$ and $q_f$ are the fermion mass and electric charge, 
$N_c$ is the number of colors, and $c_v^{(f)}$ and $c_a^{(f)}$ are the vector and axial-vector couplings to the $Z$ boson. 
In these expressions it has been assumed that $M_2-M_1\gg 2m_f$, such that the final state fermions are relativistic. 
We note that the amplitudes of the processes mediated by gauge bosons interfere with each other, but not with the amplitude of the process mediated by the Higgs, 
as the fermions in the final state have the same chirality in the former case, while opposite chirality in the latter. 
We also note that the conservation of angular momentum requires the two (relativistic) final state fermions to be emitted in the same direction when the fermions have the same chirality, 
and in opposite directions when they have opposite chirality. The conservation of linear momentum requires the scalar $\phi_1$ to be emitted collinearly with one of the fermions when they have opposite chirality
(and, when the two fermions have the same chirality, in the opposite direction to these).

The total decay rate for the decay $\phi_2\rightarrow \phi_{1}f\bar{f}$ is
\begin{align}
\Gamma_{\phi_{1}f\bar{f}}=&\frac{f^{2}_{1}N_c\alpha_{\rm EM}}{120960\pi^3\Lambda^{4}} M_2^5\Delta^5\left[252 q_f^2\, {_{2\!}F}_{1}(1,3,6;\Delta)+36 q_f c_v^{(f)}\frac{M_2^2 \Delta^2}{\cos^2\theta_{\rm W} m^{2}_{Z}} {_{2\!}F}_{1}(2,4,8;\Delta)\right.\nonumber \\
&\left.+(c_v^{(f)2}+c_a^{(f)2})\frac{M^4_2\Delta^4}{\cos^4\theta_{\rm W} m^{4}_Z} {_{2\!}F}_{1}(3,5,10;\Delta)\right]+\,\frac{f^{2}_{3}N_c m^{2}_{f}}{3840\pi^{3}m^{4}_h}M^{3}_{2}\Delta^{5}{_{2\!}F}_{1}(2,3,6;\Delta)\;,
\label{eqn:totffbarwidth}
\end{align}
where we have used that
\begin{align}
\int_0^1dx\, x^{b-1} (1-x)^{c-b-1}(1-\Delta x)^{-a}=    {_{2}F}_1(a,b,c;\Delta) {\rm B}(b,c-b)
\label{eq:hypergeometric}
\end{align}
for $c>b>0$. Here, ${\rm B}(a,b)$ is the Euler's beta function and $_2F_1(a,b,c;\Delta)$ is the Gauss's hypergeometric function, which is monotonically increasing with $\Delta$ and takes limiting values
\begin{align}
_2F_1(a,b,c,0)&=1\;, \nonumber \\
_2F_1(a,b,c,1)&=\frac{\Gamma(c) \Gamma(c-a-b)}{\Gamma(c-a)\Gamma(c-b)}\;,
\end{align}
for $c>a+b$. 

One can check that 
\begin{align}
\frac{M_2^2\Delta^2}{m_Z^2}\leq \Big(1+\frac{M_1}{M_2}\Big)^2< 4\;,
\end{align}
where the maximum value occurs for $M_2=M_1+m_Z$, namely when the $Z$-boson can be produced on-shell, and when $M_2/m_Z\rightarrow \infty$. 
Therefore, in most of the parameter space the contribution to the rate from the $Z$-boson mediated decay can be neglected against the contribution from the photon mediated decay. 
On the other hand, the contribution from the Higgs boson should not be neglected, despite the suppression by $m_f/m_h$, as it depends on a different coupling. 
A special case is the decay $\phi_2\rightarrow\phi_1\nu\bar{\nu}$, since both the Higgs and photon exchange contributions to the rate are very suppressed compared to the $Z$-boson exchange contribution. 

The process $\phi_2\rightarrow \phi_1 \gamma\gamma$, on the other hand, receives contributions from the dimension-six operator 
proportional to $f_2$ in Eq.~(\ref{eq:eff-Lag}), and from the mediation of an off-shell Higgs boson, via the dimension-four operator proportional 
to $f_3$ in Eq.~\eqref{eq:Higgs_portal} combined with the effective Higgs interaction $c_{\gamma\gamma} \frac{h}{v} F^{\mu\nu}F_{\mu\nu}$. The differential rate reads:
\begin{align}
\frac{d\Gamma_{\phi_1\gamma\gamma}}{dx_\gamma}=\frac{1}{192\pi^3}
\left(\frac{f_2}{\Lambda^2}+\frac{f_3 c_{\gamma\gamma}}{m_h^2}\right)^2
M_2^5\,\Delta^7\,\frac{x_\gamma^3(1-x_\gamma)^{3}}{(1-x_\gamma \Delta)^{3}}\;,
\label{eq:2gamma-spectrum}
\end{align}
where $x_\gamma\equiv\frac{2E_{\gamma}}{M_{2}\Delta}$ and  $c_{\gamma\gamma}\simeq -2.03 \times 10^{-3}$ in the Standard Model. Due to the conservation of angular momentum, the two photons  must be emitted back to back if they have the same polarization, and collinearly if they have opposite polarization; the conservation of linear momentum requires $\phi_1$ to be emitted along with one of the photons in the former case, and in the direction opposite to the photons in the latter.~\footnote{It is interesting to remark that, even if the photons are emitted in the same direction and with the same speed, the propagation history of the two photons on their way to the Earth might be different. Therefore they will not arrive to the detector in coincidence. We will make this assumption when we analyze the observable signals of this framework. On the other hand, the emission of two photons in exactly the same direction and with the same speed is a very peculiar feature of the decay $\phi_2\rightarrow \phi_1\gamma\gamma$, not exclusively of the framework where $\phi_1$ and $\phi_2$ are cosmologically long-lived, and could have implications in other contexts.} The partial rate of this decay channels is:
\begin{align}
\Gamma_{\phi_1\gamma\gamma}= \frac{1}{26880 \pi^3}\left(\frac{f_2}{\Lambda^2}+\frac{f_3 c_{\gamma\gamma}}{m_h^2}\right)^2M_2^5\,\Delta^7\, {_{2}F}_1(3,4,8;\Delta)\;,
\end{align}
where we have used Eq.~(\ref{eq:hypergeometric}). Here, ${_2}F_{1}(3,4,8;\Delta)$ varies between 1 and 35 for $\Delta$ between 0 and 1. 

\begin{figure}[t!]
	\begin{center}
		\includegraphics[width=.57\textwidth]{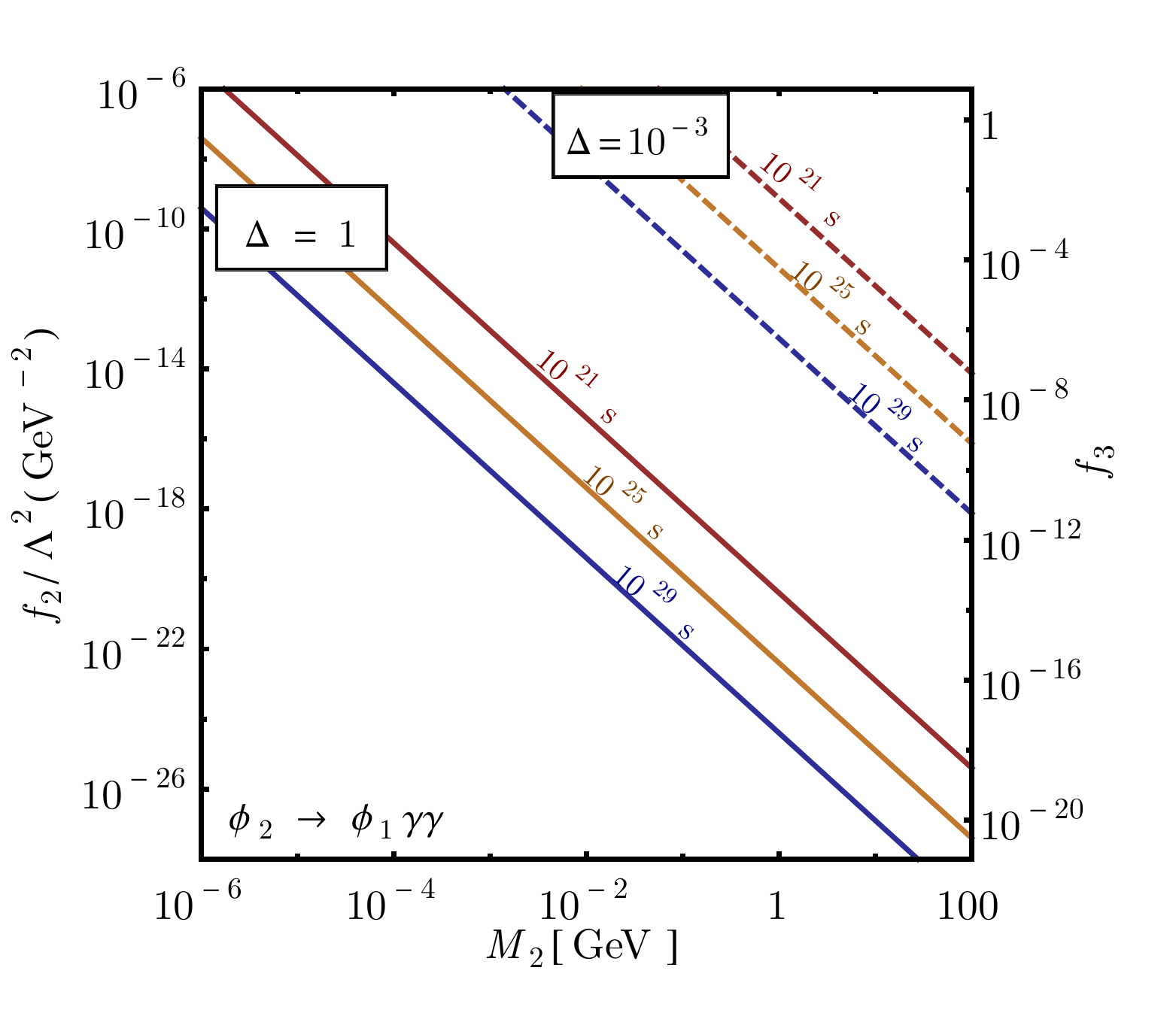}\\
		\includegraphics[width=.49\textwidth]{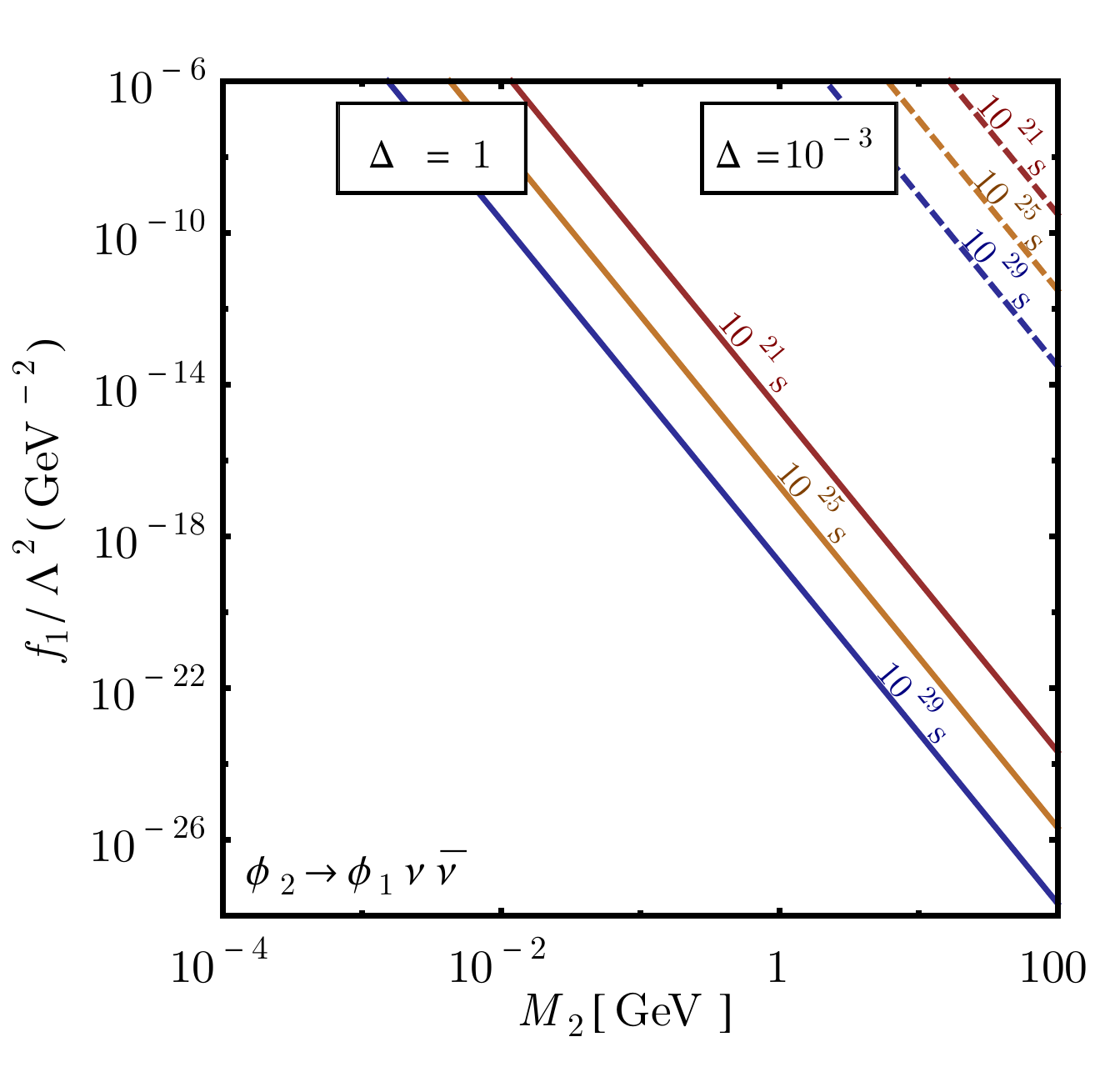}
		\includegraphics[width=.49\textwidth]{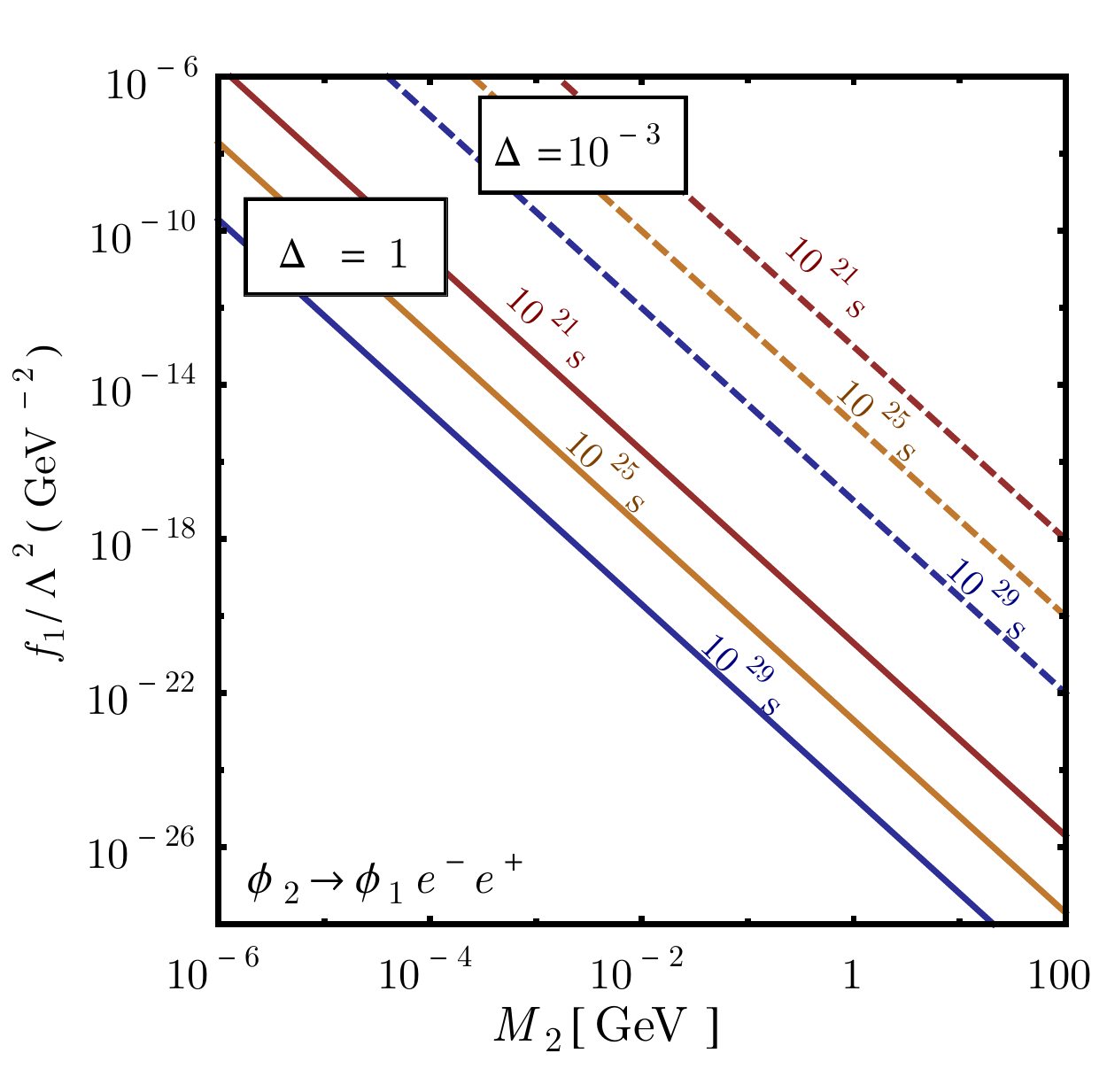}
		\caption{Inverse width contours for the decay processes $\phi_{2}\rightarrow\phi_{1}\gamma\gamma$  (top panel), $\phi_{2}\rightarrow\phi_{1}\nu\bar{\nu}$
			(bottom  left panel) and $\phi_{2}\rightarrow\phi_{1}e^{-}e^{+}$ 
			(bottom right panel) as a function of the mass of the decaying dark matter component $M_2$, 
			for exemplary hierarchical ($\Delta=1$) and degenerate ($\Delta=10^{-3}$) spectra, for the 
			effective theory described in Section \ref{sec:eff_theory} characterized by the couplings $f_1/\Lambda^2$, $f_2/\Lambda^2$ and $f_3$.}	
		\label{fig:invGamma}
	\end{center}
\end{figure}

Approximate expressions for the partial decay rates are:
\begin{align}
\Gamma_{\phi_1e^+e^-}\simeq&\Big( 10^{26} \,{\rm s}\Big)^{-1} \left[
\,\left(\frac{f_{1}/\Lambda^2}{ 1.1\times 10^{-22}\,{\rm GeV}^{2}}\right)^{2} \left(\frac{M_{2}}{1\,{\rm GeV}}\right)^{5}\Delta^5
{_{2}F}_1(1,3,6;\Delta)\right. \nonumber \\
&\left.+\left(\frac{f_3}{8.8\times 10^{-16}}\right)^{2} \left(\frac{M_{2}}{1\,{\rm GeV}}\right)^{3}\Delta^5 {_{2}F}_1(2,3,6;\Delta)
\right] \;, \nonumber\\
\Gamma_{\phi_1 \nu_i \bar \nu_i}\simeq& \displaystyle{\Big( 10^{26} \,{\rm s}\Big)^{-1}\left(\frac{f_{1}/\Lambda^2}{ 1.7\times 10^{-17}\,{\rm GeV}^{-2}}\right)^{2}
\left(\frac{M_{2}}{1\,{\rm GeV}}\right)^{9}\Delta^9{_{2}F}_1(3,5,10;\Delta)}\;,
\nonumber \\
\Gamma_{\phi_1\gamma\gamma}\simeq& \Big( 10^{26} \,{\rm s}\Big)^{-1}\,\left[\frac{f_{2}/\Lambda^2}{ 7.4\times 10^{-23}\,{\rm GeV}^{-2}}
-\frac{f_3}{5.7\times 10^{-16}}\right]^2\left(\frac{M_{2}}{1\,{\rm GeV}}\right)^{5}\Delta^7 {_{2}F}_1(3,4,8;\Delta).
\label{eq:approx_rates}
\end{align}
Clearly, $\phi_2$ can be cosmologically long-lived for sufficiently weak interaction strengths $f_1$, $f_2$, $f_3$, and/or for a small mass for the mother dark matter particle and/or for a small mass difference with the daughter dark matter particle. Fig.~\ref{fig:invGamma} shows contour lines of the inverse width in the final states $\phi_1\gamma\gamma$ (top panel), $\phi_1\nu\bar\nu$ (bottom panel left) and $\phi_1 e^- e^+$ (bottom panel right), for the representative cases $\Delta=1$ and $\Delta=0.001$, which respectively correspond to a very hierarchical spectrum and to a very degenerate spectrum of dark matter components. 

\begin{figure}[t!]
	\begin{center}
		\includegraphics[width=.49\textwidth]{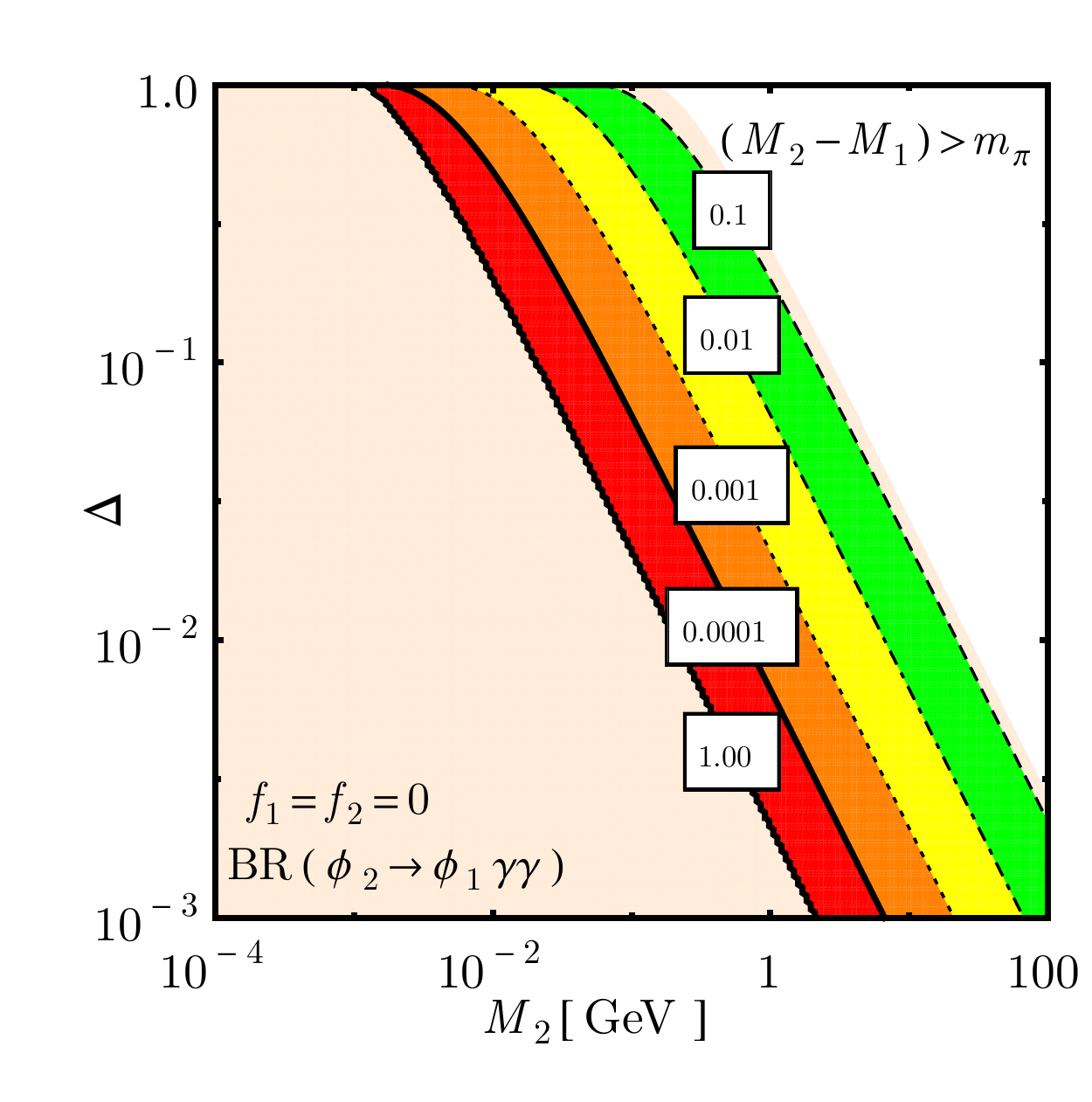}
		\includegraphics[width=.49\textwidth]{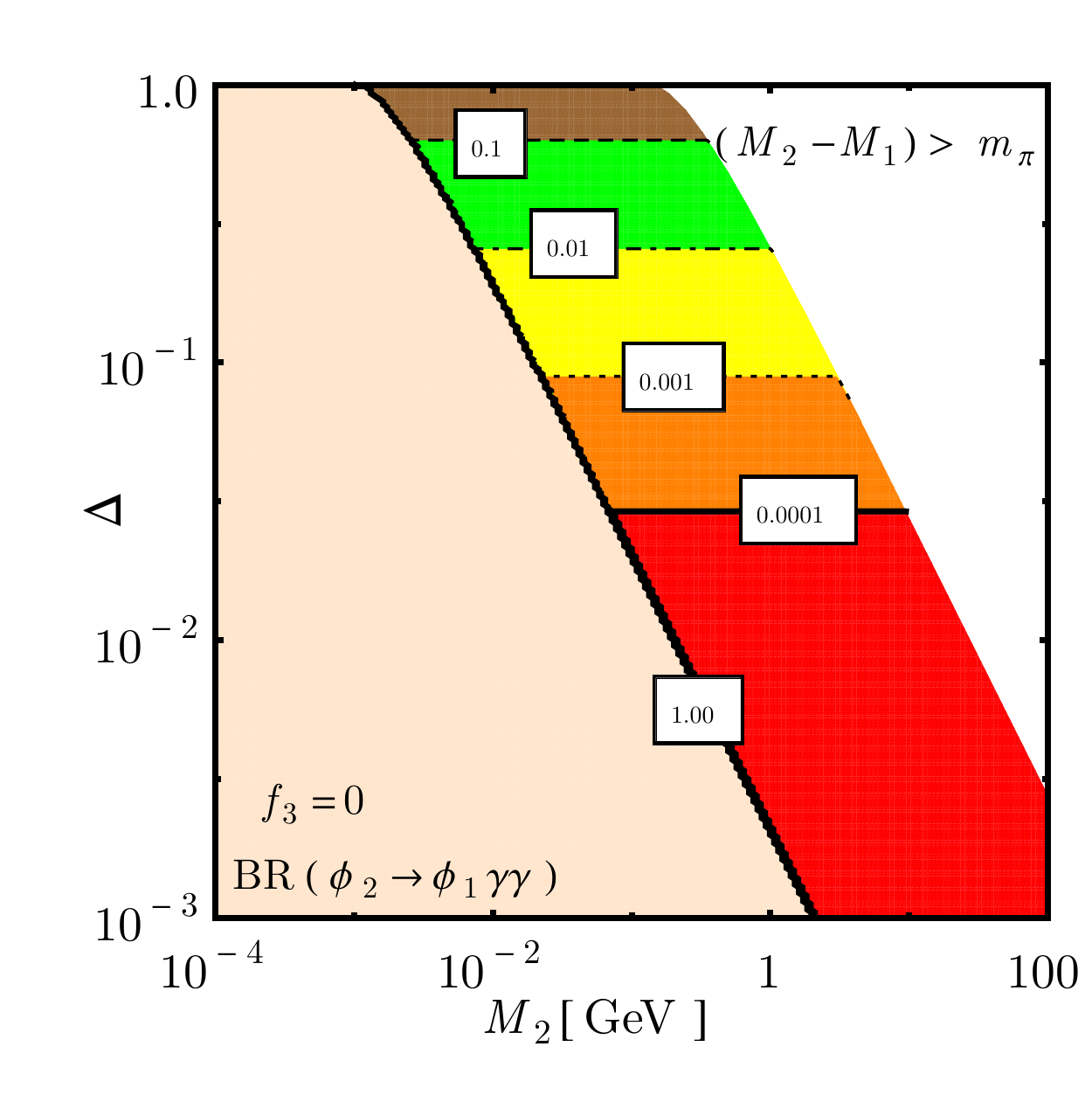}	
		\caption{Branching ratios of the process  $\phi_{2}\rightarrow\phi_{1}\gamma\gamma$ as a function of the decaying dark matter mass $M_2$ and the degeneracy parameter $\Delta$ for an effective theory with $f_1=f_2=0$ (left panel) and with $f_3=0$ (right panel). The white region, where the decay $\phi_2\rightarrow \phi_1 \pi^0$ is kinematically accessible, is disregarded in our analysis. See main text for details. }
		\label{fig:BRs}
	\end{center}
\end{figure}

The rates for the different processes depend on different combinations of the couplings $f_1$, $f_2$ and $f_3$. However, if the decays are dominated by the Higgs exchange one finds
\begin{align}
\label{eqn:h_med_ratio}
\frac{\Gamma_{\phi_1\gamma\gamma}}{\Gamma_{\phi_1 e^- e^+}}&\simeq 
\frac{c^{2}_{\gamma\gamma}}{7}\frac{M_2^2 \Delta^2}{m_f^2}\,\frac{{_{2}F}_1(3,4,8;\Delta)}{{_{2\!}F}_{1}(2,3,6;\Delta)}\;, \nonumber\\
\Gamma_{\phi_1 \nu_i \bar \nu_i}&\simeq 0\;.
\end{align}
Besides, for the toy model where the effective interactions $f_1$ and $f_2$ are generated via integrating out a vector-like pair of heavy fermions,
\begin{align}
\label{eqn:eff_opp_ratio}
\frac{\Gamma_{\phi_1\gamma\gamma}}{\Gamma_{\phi_1e^+e^-}}&\simeq \,\frac{\pi}{56}\,\frac{\Delta^2\,g^{-2}\left(m_1/m_2\right)}{ \tan^2\arg(Y_1 Y_2^*)} \frac{{_{2}F}_1(3,4,8;\Delta)}{{_{2\!}F}_{1}(1,3,6;\Delta)}\;,\nonumber\\
\frac{\Gamma_{\phi_1\gamma\gamma}}{\sum_i\Gamma_{\phi_1 \nu_i \bar \nu_i}}&\simeq \frac{3\pi}{2\,\left(c^{(f)\,2}_{v}+c^{(f)\,2}_{a}\right)}\,\frac{g^{-2}\left(m_1/m_2\right)}{\tan^2\arg(Y_1 Y_2^*)} \,\frac{m^{4}_{z}\,\cos^{4}\theta_{W}}{M^{4}_{2}\Delta^{2}}\,\frac{\,{_{2}F}_1(3,4,8;\Delta)}{{_{2\!}F}_{1}(3,5,10;\Delta)}\;.
\end{align}
The branching ratios for the decay $\phi_{2}\rightarrow\phi_{1}\gamma\gamma$ both for the Higgs mediated scenario (left) 
and for the fermion loop scenario (right) are shown in Fig.~\ref{fig:BRs} for different values of $\Delta$ with $M_{2}-M_{1}\,\leq\,m_{\pi}$, 
taking for concreteness $m_2 \simeq 5\,m_1$ and $\arg(Y_1 Y_2^*)=\pi/4$. One concludes from the plot that the decay $\phi_2\rightarrow\phi_1\gamma\gamma$	
has a sizable or dominant branching ratio in a substantial part of the parameter space. Therefore, the model could be testable with gamma-ray observations.

\begin{figure}[t!]
	\begin{center}
		\includegraphics[width=0.49\textwidth]{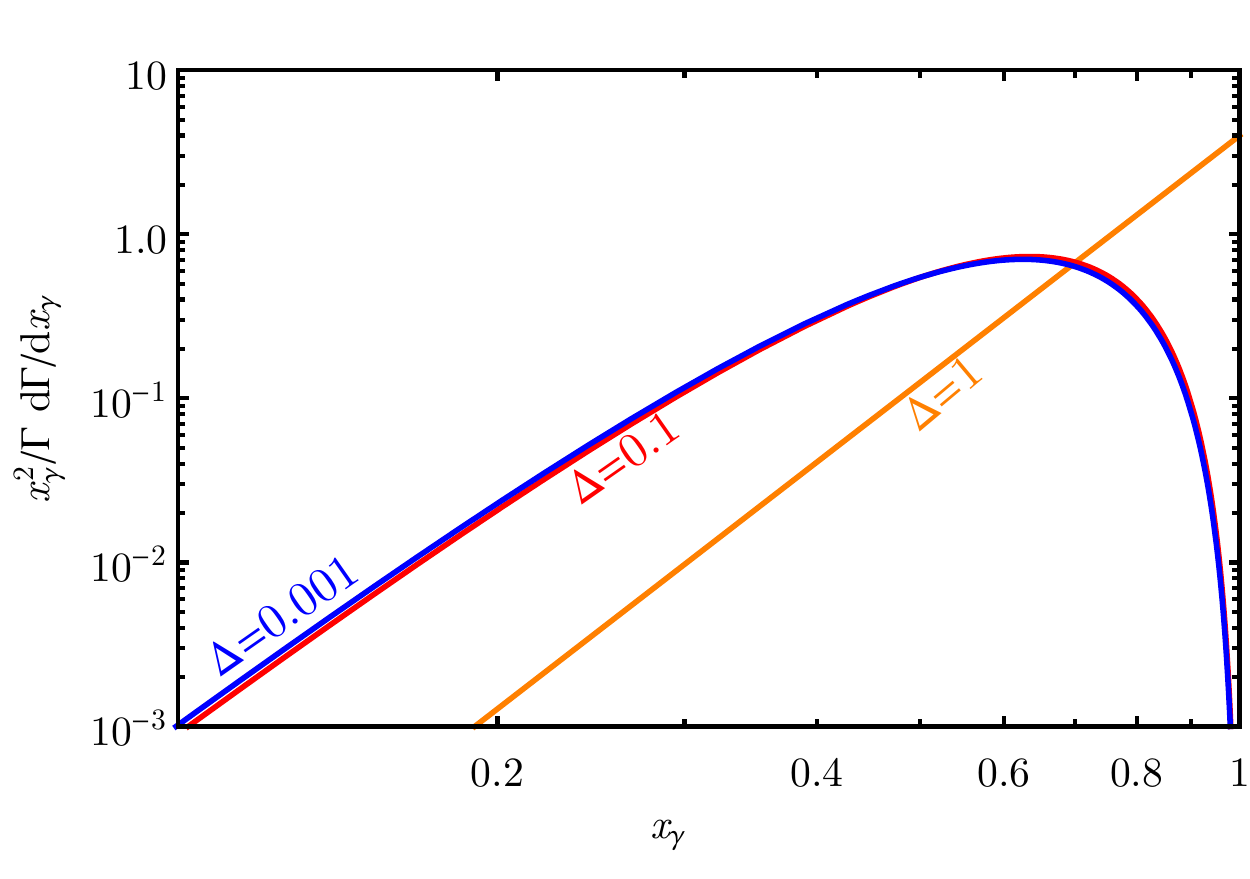}\\\vspace{1mm}
		\caption{Differential energy spectrum of photons produced in the decay $\phi_2\rightarrow \phi_1\gamma\gamma$ for different values of the degeneracy parameter $\Delta$ when the decay process can be described by the effective theory presented in Section \ref{sec:eff_theory}.}
		\label{fig:diff_spectrum_eff}
	\end{center}
\end{figure}

The (normalized) differential energy spectrum in this scenario can be cast as:
\begin{align}
\frac{1}{\Gamma_{\phi_1\gamma\gamma}}\frac{d\Gamma_{\phi_1\gamma\gamma}}{dx_{\gamma}}=\frac{140}{{_{2}F}_1(3,4,8;\Delta)}\,\frac{x_{\gamma}^3(1-x_{\gamma})^{3}}{(1-x_{\gamma}\,\Delta)^{3}}\;,
\label{eqn:2gamma-spectrum_eff_norm}
\end{align}
which only depends on the mass splitting parameter $\Delta$ and on the variable $x$. The energy spectrum is shown in Fig.~\ref{fig:diff_spectrum_eff}, 
for three representative values of $\Delta$. The spectrum presents a maximum at $x= (1-\sqrt{1-\Delta})/\Delta$, which takes values between 1/2 and 1 
for $\Delta\rightarrow 0$ and $\Delta\rightarrow 1$. In terms of the photon energy, this is equivalent to a peak at $E_\gamma=\frac{M_2}{2}(1-\sqrt{1-\Delta})$, 
which takes values between $E_\gamma=\frac{1}{4} M_2\Delta$ and $E_\gamma=\frac{1}{2} M_2\Delta$ for $\Delta\rightarrow 0$ and  $\Delta\rightarrow 1$, respectively. 
Notably, and regardless of the value of the parameter $\Delta$, the energy spectrum presents a sharp peak close to the kinematical endpoint, which could stand out over 
the featureless spectrum of the isotropic diffuse gamma-ray emissions.

The flux of photons received on Earth from the decay $\phi_2\rightarrow\phi_1\gamma\gamma $ receives two main contributions.
The extragalactic contribution is generated by the decay of dark matter particles distributed homogeneously and isotropically 
in the Universe, and has a differential spectrum given by
\begin{align}
\frac{d\Phi_\text{eg}}{dE_\gamma} =\frac{1}{4\pi}\frac{\Omega_{\phi_2}
	\rho_{\rm c} }{M_2}
\int_0^\infty \frac{dz}{H(z)} \frac{d\Gamma}{dE_\gamma} \left[(z+1)E_\gamma\right]
\;e^{-\tau(E_\gamma,z)}\;,
\label{eq:gamma-dec-EG}
\end{align}
which includes contributions from all redshifts $z$. Here, $\rho_{\rm c}=4.9\times10^{-6}{\rm GeV}{\rm cm}^{-3}$ is the critical
density of the Universe, $H(z)=H_0 \sqrt{\Omega_\Lambda+\Omega_\text{m}(z+1)^3}$ is the (redshift-dependent) Hubble parameter, 
and $\tau(E_\gamma,z)$ is the optical depth, which determines the attenuation of the gamma-ray flux in their propagation from the 
decay point to the Earth. In our analysis we have adopted $\Omega_\Lambda=0.69$, $\Omega_{\rm m}=0.31$ \cite{Ade:2015xua}, and the 
parametrization of the optical depth presented in \cite{Cirelli:2010xx}. 
The second contribution stems from the decay of dark matter particles in the Milky Way halo, and is given by:
\begin{equation}
\frac{d\Phi_\text{halo}}{dE_\gamma}(\psi) =\frac{1}{4\pi M_2} \frac{\Omega_{\phi_2}}{\Omega_{\rm DM}}
\frac{d\Gamma}{dE_\gamma} 
\int_0^\infty ds\, \rho_{\rm DM}[r(s,\psi)]\;, 
\label{eq:flux-gamma-decay}
  \end{equation}
where we have assumed that the fraction of the dark matter mass density 
in the form of the unstable component $\phi_2$ is the same in the Milky 
Way and in the Universe at large scale: $\rho_{\phi_2}/\rho_{\rm DM}=
\Omega_{\phi_2}/\Omega_{\rm DM}$. In determining the galactic
contribution to the gamma-ray flux, $\dfrac{d\Phi_\text{halo}}{dE_\gamma}
(\psi)$ we have assumed the Navarro-Frenk-White (NFW) dark matter halo 
profile~\cite{Navarro:1996gj} with scale radius $r_{s}\,=\,21\,$kpc,  
local dark matter density  $\rho_{\odot}\,=\,0.3\,\rm{GeV/cm^{3}}$ and distance of the Sun to the Milky Way center  $r_{\odot}\,=\,8.5\,$kpc.

\begin{figure}[t!]
	\begin{center}
		\includegraphics[width=0.49\textwidth]{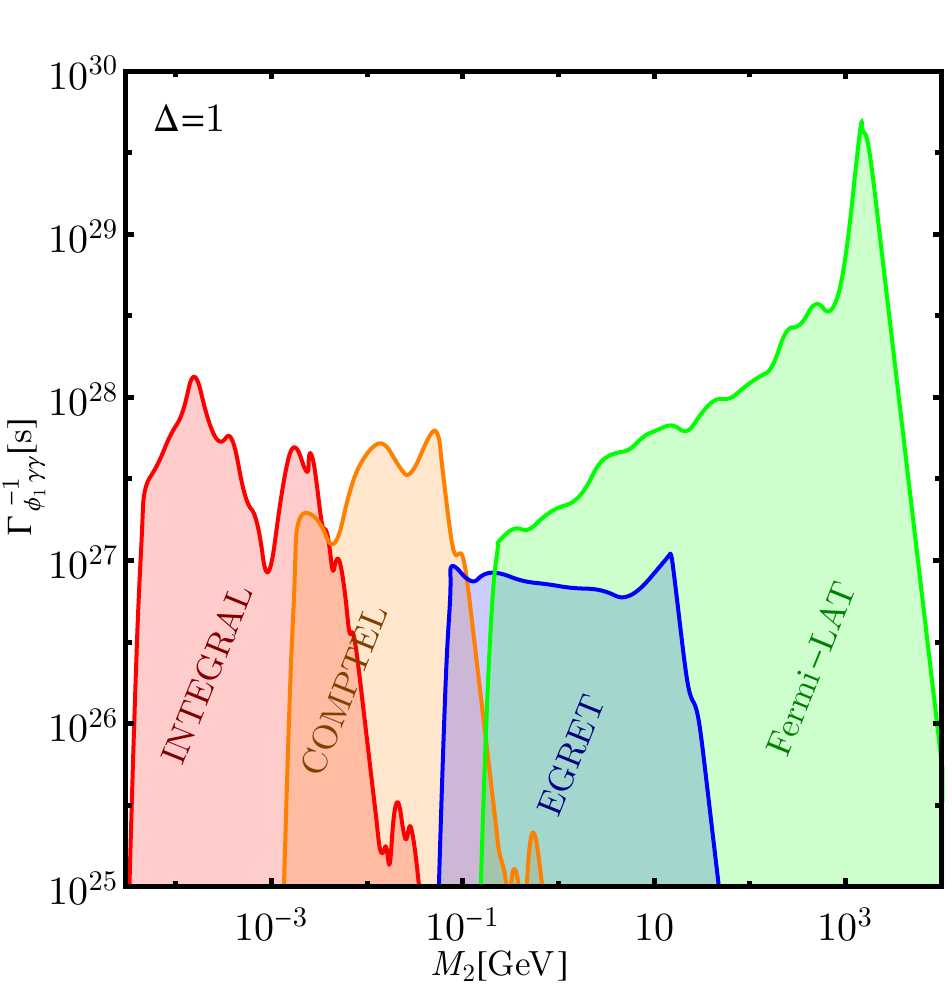}
		\includegraphics[width=0.49\textwidth]{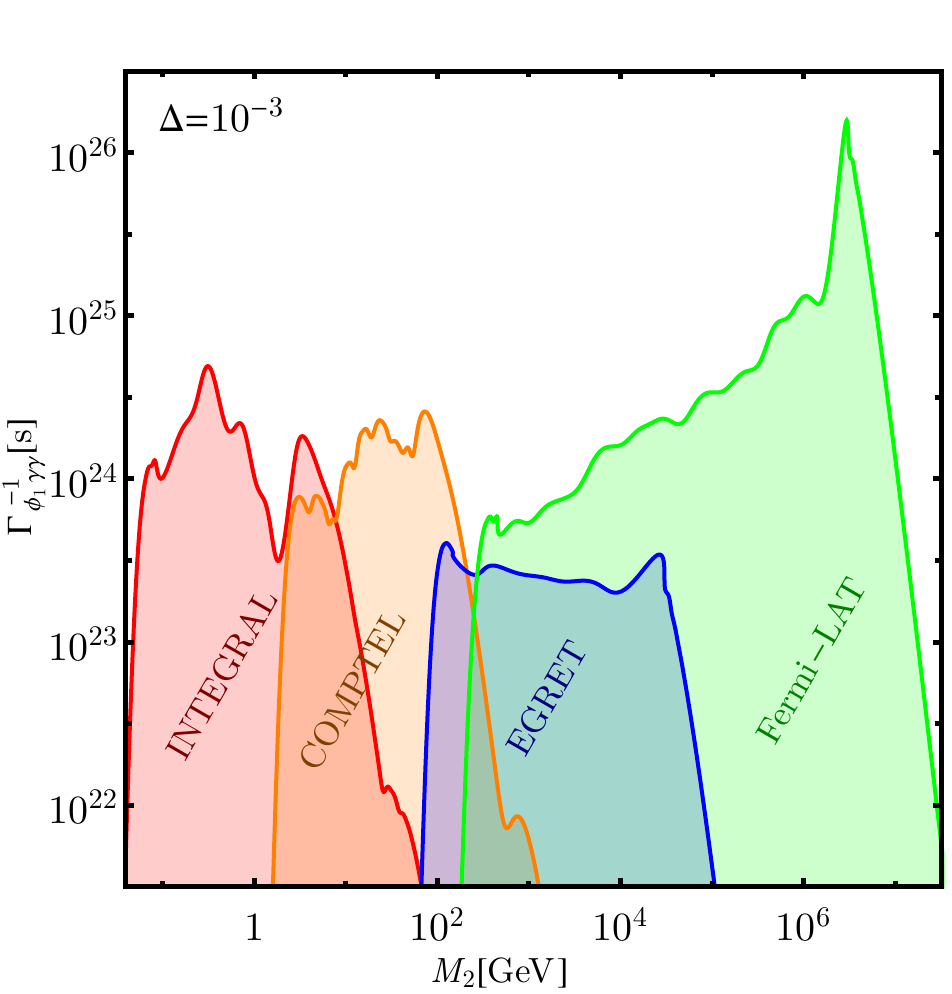}\\\vspace{1mm}
		\caption{Lower limit on the inverse width of the decay process $\phi_{2}\rightarrow\phi_{1}\gamma\gamma$ as a function of the mass of the decaying dark matter component $M_2$, for a very hierarchical ($\Delta=1$, left panel) and a very degenerate ($\Delta=10^{-3}$, right panel) dark matter mass spectrum. }
		\label{fig:lifetime_limit}
	\end{center}
\end{figure}

The non-observation of a statistically significant sharp feature in the 
isotropic diffuse photon flux determined by INTEGRAL~\cite{Bouchet:2008rp}, COMPTEL~\cite{1999ApL&C..39..193W}, EGRET~\cite{Strong:2004de} and 
the Fermi-LAT~\cite{Ackermann:2014usa} leads to limits on the width of the dark matter decay channel $\phi_2\rightarrow \phi_1 \gamma\gamma$, 
which are shown in Fig.~\ref{fig:lifetime_limit} as a function of the dark matter mass, and for two different values of the mass splitting parameter 
$\Delta=1$(left panel) and $10^{-3}$(right panel). The limits have been derived assuming that $\Omega_{\phi_2}h^2=\Omega_{DM}h^2$ and imposing that 
the predicted photon flux does not exceed the $2\sigma$ limit reported by the experiment in every energy bin. We have assumed a flat energy resolution $\sigma(E)/E=0.1$ throughout our analysis.
The contribution to the photon flux from inverse Compton scattering has been neglected; therefore our limits can be regarded as conservative. 
In principle, a given experiment is sensitive to arbitrarily large dark matter masses, due to the low energy tail in the photon energy spectrum. 
On the other hand, the photon multiplicity decreases rapidly, as $x_\gamma^{-3}$, when $x_\gamma\ll 1$, while the measured photon flux scales roughly as $E_\gamma^{-2}$. 
As a result, the range of masses to which the experiment is sensitive is effectively bounded from above as well. 
	
	We obtain that current observations require $\Gamma^{-1}_{\phi_1\gamma\gamma}\gtrsim 5\times10^{26}$ s for $\Delta=1$ and for a mother particle with mass $M_2$ in the range 40 keV $-$ 1 TeV. 
	As the spectrum becomes more and more degenerate, the limits on the inverse width become weaker, approximately by a factor $\Delta/2$ (this is due to the fact that the energy of the peak is 
	proportional to $1-\sqrt{1-\Delta}\simeq \Delta/2\,$, combined with the fact that the observational limits are roughly flat with the energy of the photon). For $\Delta=10^{-3}$ 
	the inverse width is therefore restricted to be $\Gamma^{-1}_{\phi_1\gamma\gamma}\gtrsim 2.5\times 10^{23}$ s when the mother dark matter particle mass is in the range $M_2=80\,{\rm MeV}-2\,{\rm PeV}$.		

\section{Scalar dark sector coupled to Standard Model fermions}
\label{sec:coupling-to-SM}

In the previous section we have considered a possible UV completion to the effective interaction Eq.\eqref{eq:eff-Lag} consisting  in one heavy $Z_2$-even fermion 
and one heavy $Z_2$-odd fermion, with the same gauge quantum numbers. The Standard Model contains already various $Z_2$-even fermions, therefore, an obvious variant 
of the aforementioned scenario consists in identifying $\psi_1$ with any Standard Model fermion, which we denote by $f$, and $\psi_2$ with a heavy $Z_2$-odd exotic fermion. 
If the Standard model fermion is also heavy, $m_1\gg M_2,M_1$, the results of the previous section apply. However, if the dark matter particles interact with a light Standard Model fermion, 
a separate analysis is necessary. In this section we focus in the scenario where $ M_2-M_1\gg m_1$, so that the decay $\phi_2\rightarrow \phi_1 f\bar f$ proceeds at tree level (since $\psi_1$ is identified with $f$). 
The decay $\phi_2\rightarrow \phi_1 \gamma\gamma$, on the other hand, still proceeds at the one loop level. However, due to the lightness of the Standard Model fermion in the loop, 
the process cannot be described by the effective interactions constructed in the previous Section.

 For simplicity, we will assume in this section that the Higgs portal interaction is negligible, so that the decays proceed dominantly by the interactions with the Standard Model 
 fermion and the $Z_2$-odd exotic fermion.
The amplitude for the process $\phi_2\rightarrow \phi_1 f \bar f$ can be obtained from the effective interaction 
\begin{eqnarray}
-\mathcal{L}_{int}\,\supset\,\frac{{\rm Re}(Y_1 Y_2^*)}{2m^{2}_{2}}\,\bar{f}\gamma^{\mu}\left(a\,P_{L}+b\,P_{R}\right)f\,\left(\phi_{1}\overset{\leftrightarrow}{\partial_{\mu}}\,\phi_{2}\right),
\label{eqn:ffbar_effop}
\end{eqnarray}
where $P_{L,R}$ are the chiral projection operators. On the other hand, and as said above, the decay $\phi_2\rightarrow \phi_1 \gamma\gamma$ cannot be described by an effective interaction and the amplitude must be calculated instead from the full Lagrangian. We obtain
\begin{equation}
\mathcal{A}(\phi_{2}\rightarrow\phi_{1}\gamma\gamma)=\frac{\alpha_{\rm EM}}{\pi}\frac{m_1}{m_2}{\rm Re}(Y_1 Y_2^*)\,{\cal I}\left(\frac{m_1^2}{k_{1}k_{2}}\right)\,\left[\eta^{\mu\nu}-\frac{k_{1}^\mu\,k_{2}^\nu}{k_1 k_2}\right]\epsilon_{1\nu} \epsilon_{2\mu}\;,
\label{eqn:digam_amp}
\end{equation}
where $k_{1,2}$ and $\epsilon_{1,2}$ are the four-momenta and the polarizations of the emitted photons, and the loop function ${\cal I}(x)$ is given by
\begin{eqnarray}
{\cal I}(x)\,&=&1+\frac{1}{2}(1-2x)\Big[{\rm Li}_2\left(\frac{1+\sqrt{1-2x}}{x}\right)+{\rm Li}_2\left(\frac{1-\sqrt{1-2x}}{x}\right)\Big]\;.
\label{eqn:loop_func}
\end{eqnarray}
For $x\geq\, 1/2$ the function ${\cal I}(x)$ is real and monotonically decreasing, while for $x\,\leq\,1/2$, ${\cal I}(x)$ contains an imaginary part due to 
the on-shellness of the loop fermion $\psi_{1}$. Approximate expressions for ${\cal I}(x)$ are
\begin{eqnarray}
{\cal I}(x)\,&\approx&\begin{cases}
\displaystyle{\left(1+\frac{\pi^2}{4}\right)
	-\frac{1}{4}(1-2x)\log^2\frac{x}{2} -\frac{x}{2}(\pi^2+\log\frac{x}{2})+i\frac{\pi}{2} (1-2x)\log\frac{x}{2}}&~~{\rm for}\,x\ll 1\\\nonumber
\displaystyle{\frac{1}{3x}+\frac{7}{180x^{2}}}&~~{\rm for}\,x\gg 1
\end{cases}\;.\\
\label{eq:loopfunc_approx}
\end{eqnarray}

The differential rate for the process $\phi_2\rightarrow \phi_1 f \bar f$ reads, under the assumption $M_2-M_1\gg 2m_f$, 
\begin{equation}
\frac{d\Gamma_{\phi_{1}f\bar{f}}}{dx_{f}}\,=\,\frac{1}{64\pi^{3}}{\rm Re}(Y_1 Y_2^*)^2\left(\frac{M_2^5\Delta^5}{m^4_{2}}\right)\,\frac{x_f^2(1-x_f)^{2}}{(1-x_f\Delta)}\;,
\label{eqn:ffbar-spectrum_lowm1}
\end{equation}
resulting in a partial decay width
\begin{equation}
\Gamma_{\phi_{1}f\bar{f}}\,=\,\frac{1}{1920\pi^{3}}{\rm Re}(Y_1 Y_2^*)^2\left(\frac{M_2^5\Delta^5}{m^4_{2}}\right)\,{_{2}F}_1(1,3,6;\Delta).
\label{eqn:ffbar-width_lowm1}
\end{equation}

On the other hand, the differential decay width for $\phi_{2}\rightarrow\phi_{1}\gamma\gamma$ reads, 
\begin{equation}
\frac{d\Gamma_{\phi_1\gamma\gamma}}{dx_\gamma}=\frac{\alpha^2_{\rm EM}}{128\pi^5} {\rm Re}(Y_1 Y_2^*)^2\left(\frac{m^{4}_{1} \Delta}{M_{2} m^2_2}\right)\,	F_{\gamma\gamma}\left(x_{\gamma}\right)\;,
\label{eqn:2gamma-spectrum_lowm1}
\end{equation}
where
\begin{align}
F_{\gamma\gamma}=\int_{z_{\rm min}}^\infty dz \frac{|{\cal I}(z)|^{2}}{z^{2}}\approx 
\begin{cases}
\displaystyle{\frac{1}{27 z_{\rm min}^3}}&~~{\rm for}\, z_{\rm min}\gg 1
\\
\displaystyle{\frac{1}{z_{\rm min}}\left( 12.8+0.062 \log^4z_{\rm min}\right)}&~~{\rm for}\, z_{\rm min}\ll 1
\end{cases}\;.
\label{eqn:Fgg_expression}
\end{align}
with $z_{\rm min}\,=\,\dfrac{2m^{2}_{1}}{M^{2}_{2}}\dfrac{\left(1-x_{\gamma}\Delta\right)}{\left(1-x_{\gamma}\right)x_{\gamma}\Delta^{2}}$. We note that the regime $M_2-M_1< 2 m_1$ ($>2m_1$) corresponds to $z_{\rm min}>1/2$ ($<1/2$).

Thus, from Eq.~\eqref{eqn:2gamma-spectrum_lowm1} and 
\eqref{eqn:Fgg_expression} one obtains:
\begin{align}
\frac{d\Gamma_{\phi_1\gamma\gamma}}{dx_\gamma}\approx\frac{\alpha^2_{\rm EM}}{128\pi^5} {\rm Re}(Y_1 Y_2^*)^2\left(\frac{m^{4}_{1} \Delta}{M_{2} m^2_2}\right)\,\begin{cases}
\displaystyle{\frac{M^{6}_{2}\Delta^{6}}{216\,m^{6}_{1}}\frac{x^3_{\gamma}\left(1-x_{\gamma}\right)^3}{\left(1-x_{\gamma}\Delta\right)^3}}\hspace{3cm}{\rm for}\, M_{2}-M_{1}\ll 2\,m_1
\\
\displaystyle{\frac{M^2_{2}\Delta^2}{2\,m^{2}_{1}}\frac{x_{\gamma}\left(1-x_{\gamma}\right)}{\left(1-x_{\gamma}\Delta\right)}\left[ 12.8+ 0.062\log^4\left(\dfrac{2m^{2}_{1}}{M^{2}_{2}}\dfrac{\left(1-x_{\gamma}\Delta\right)}{\left(1-x_{\gamma}\right)x_{\gamma}\Delta^{2}}\right)\right]} \\
\hspace{6.5cm}\,{\rm for}\, M_{2}-M_{1}\gg 2\,m_1
\end{cases}\;.
\label{eqn:2gamma-spectrum_lowm1new}
\end{align}

Approximate expressions for the partial decay rates are:
\begin{align}
\Gamma_{\phi_1 f\bar f}&\simeq  \left(10^{26}{\rm s}\right)^{-1} \left(\frac{{\rm Re}(Y_{1}Y^{*}_{2})}{2.1\times\,10^{-17}}\right)^2 \left(\frac{m_2}{1\,{\rm TeV}}\right)^{-4} \left(\dfrac{M_{2}}{1\,{\rm GeV}}\right)^{5}\Delta^{5}{}_{2}F_{1}\left(1,3,6;\Delta\right), \nonumber
\\
\Gamma_{\phi_1\gamma\gamma} &\simeq \left(10^{26}{\rm s}\right)^{-1} \left(\dfrac{m_{2}}{1\,{\rm TeV}}\right)^{-2} \nonumber\\
&\times
\begin{cases}
\displaystyle{\left(\frac{{\rm Re}(Y_{1}Y^{*}_{2})}{6.3\times\,10^{-20}}\right)^2 \left(\frac{m_{e}}{m_{1}}\right)^{2}\left(\dfrac{M_{2}}{1\,{\rm GeV}}\right)^{5}\Delta^{7}{}_{2}F_{1}\left(3,4,8;\Delta\right)}
\hspace{1cm}{\rm for\,}M_{2}-M_{1}\,\ll\,2m_{1}\\
\!\begin{aligned}
&\displaystyle{\left(\frac{{\rm Re}(Y_{1}Y^{*}_{2})}{6\times\,10^{-15}}\right)^2 \left(\frac{m_{1}}{m_{e}}\right)^{2}\left(\dfrac{M_{2}}{1\,{\rm GeV}}\right)\Delta^{3}\times}\\\nonumber
&~~~~\Big[\left(1+\frac{1}{200}\log^4\left(\frac{M^{2}_{2}\Delta^{2}}{2m^{2}_{1}}\right)\right){}_{2}F_{1}\left(1,2,4;\Delta\right)
+0.03\,{\cal J}\left(M_{2},m_{1},\Delta\right)\Big]
\end{aligned}
\nonumber\\
\end{cases}\;,\nonumber\\
&\hspace{11.4cm}\,{\rm for\,}M_{2}-M_{1}\,\gg\,2m_{1}
\label{eq:approx_rates_SM4}
\end{align}
where, 
\begin{align}
{\cal J}\left(M_{2},m_{1},\Delta\right)\equiv& \int_0^1 dx_{\gamma}\frac{x_{\gamma}(1-x_{\gamma})}{(1-\Delta\,x_{\gamma})}\nonumber\\
&\bigg[4\log^3\left(\frac{M^{2}_{2}\Delta^2}{2m^{2}_{1}}\right)\log\left[\frac{x_{\gamma}(1-x_{\gamma})}{(1-\Delta\,x_{\gamma})}\right]+6\log^2\left(\frac{M^{2}_{2}\Delta^2}{2m^{2}_{1}}\right)\log^2\left[\frac{x_{\gamma}(1-x_{\gamma})}{(1-\Delta\,x_{\gamma})}\right]\bigg]\;.\nonumber\\
\end{align}

\begin{figure}[t!]
	\begin{center}
		\includegraphics[width=0.49\textwidth]{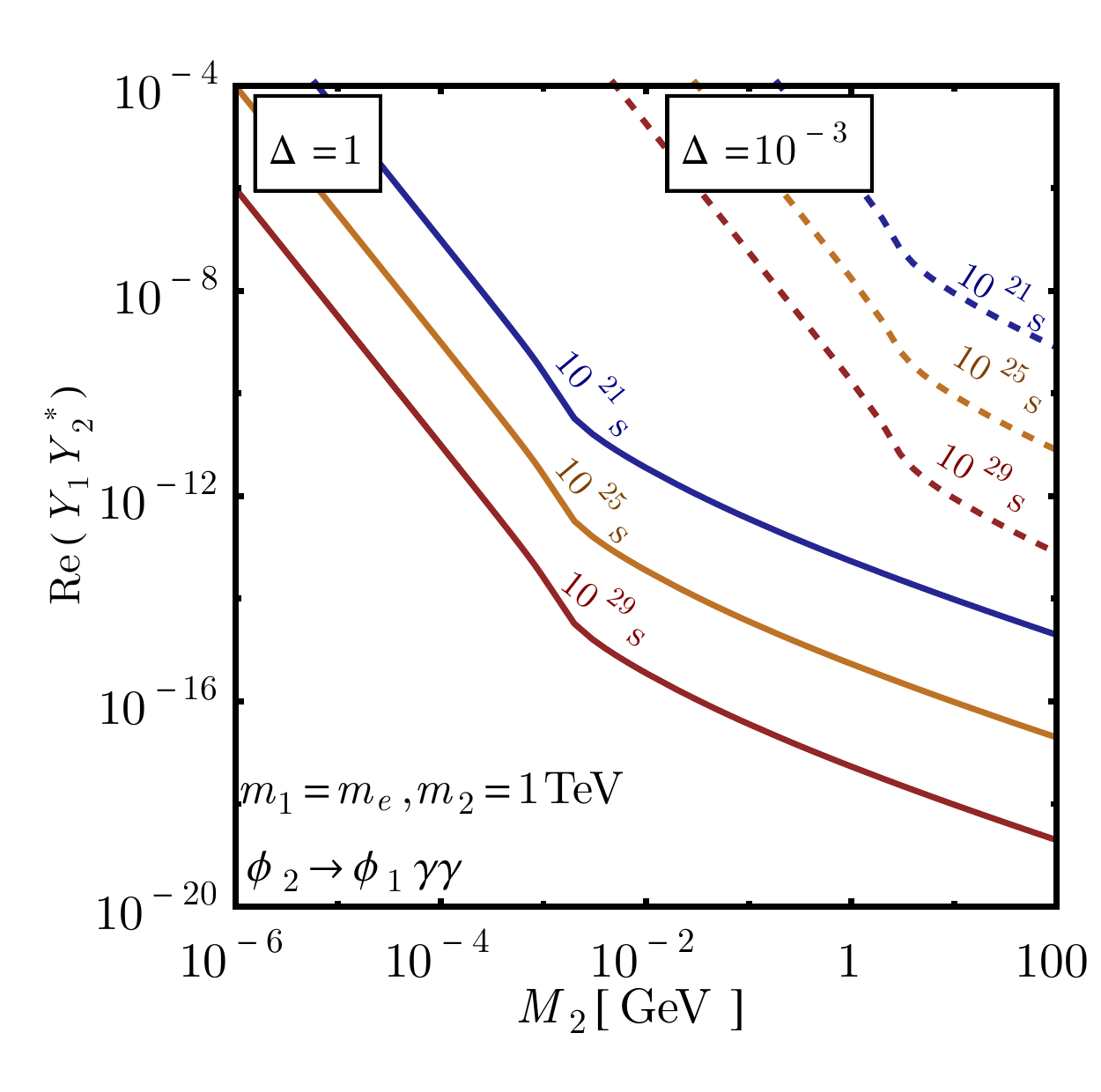}
		\includegraphics[width=0.49\textwidth]{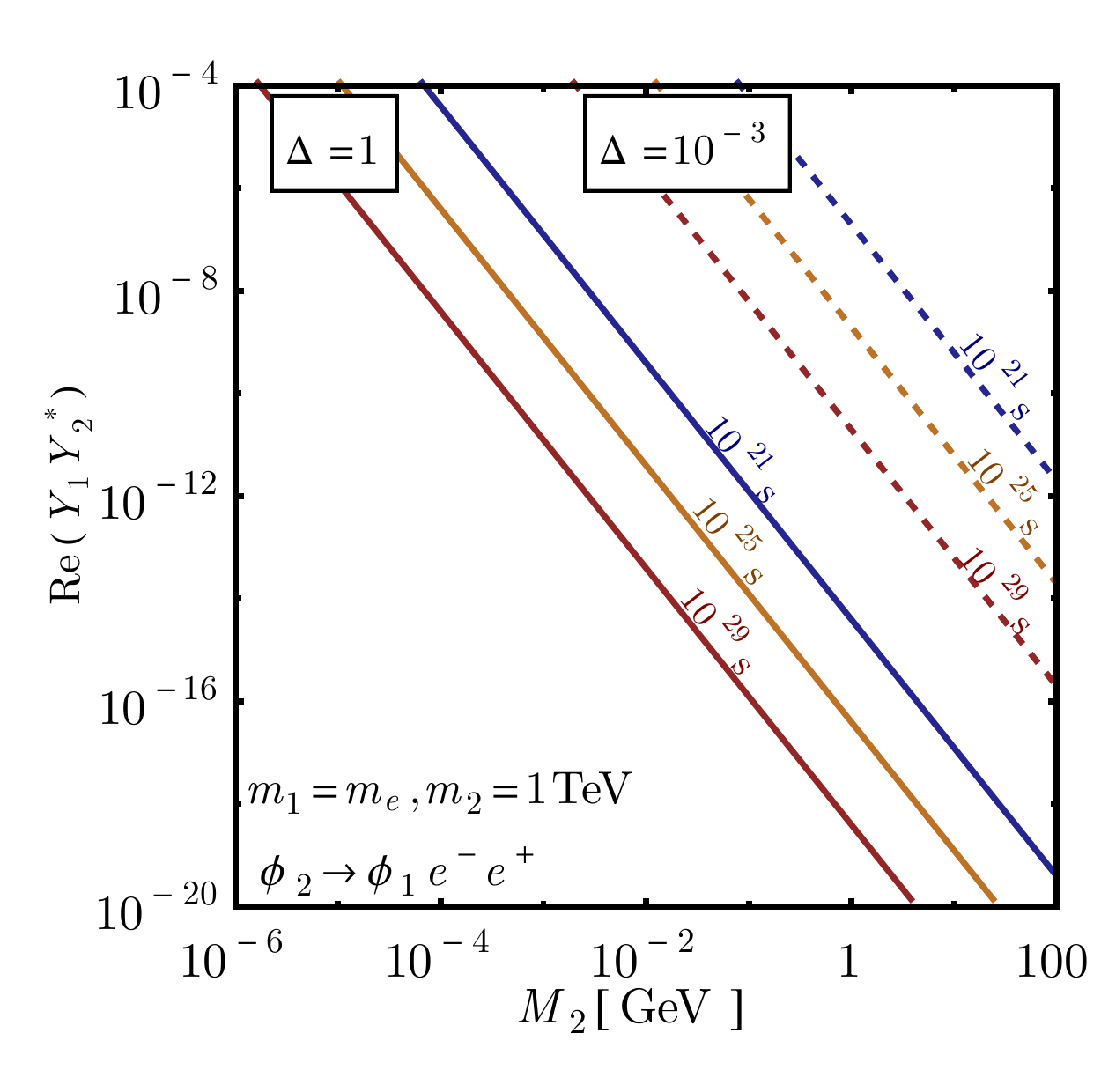}
			\caption{Same as Fig.~\ref{fig:invGamma}, but for the scenario described in Section \ref{sec:coupling-to-SM} where the two dark matter components have Yukawa couplings 
			$Y_{1,2}$ to a light $Z_2$-even fermion and a heavy $Z_2$-odd exotic fermion. For the plot we have assumed that the $Z_2$-even fermion is an electron and that the $Z_2$-odd 
			fermion has mass $m_2=1$ TeV.}
			\label{fig:invGamma_lowm1}
	\end{center}
\end{figure}

Fig.~\ref{fig:invGamma_lowm1} shows contour lines of the inverse widths into $\phi_1 \gamma\gamma$ (left panel) and into $\phi_1 e^+e^-$ (right panel), in the parameter space spanned by ${\rm Re}(Y_1 Y_2^*)$ and $M_2$, for $\Delta=1$ and $\Delta=10^{-3}$,  taking for concreteness a model where the dark matter particles couple to an electron and to a $Z_2$-odd exotic fermion with mass $m_2=1$ TeV.

\begin{figure}[t!]
	\begin{center}
		\includegraphics[width=.49\textwidth]{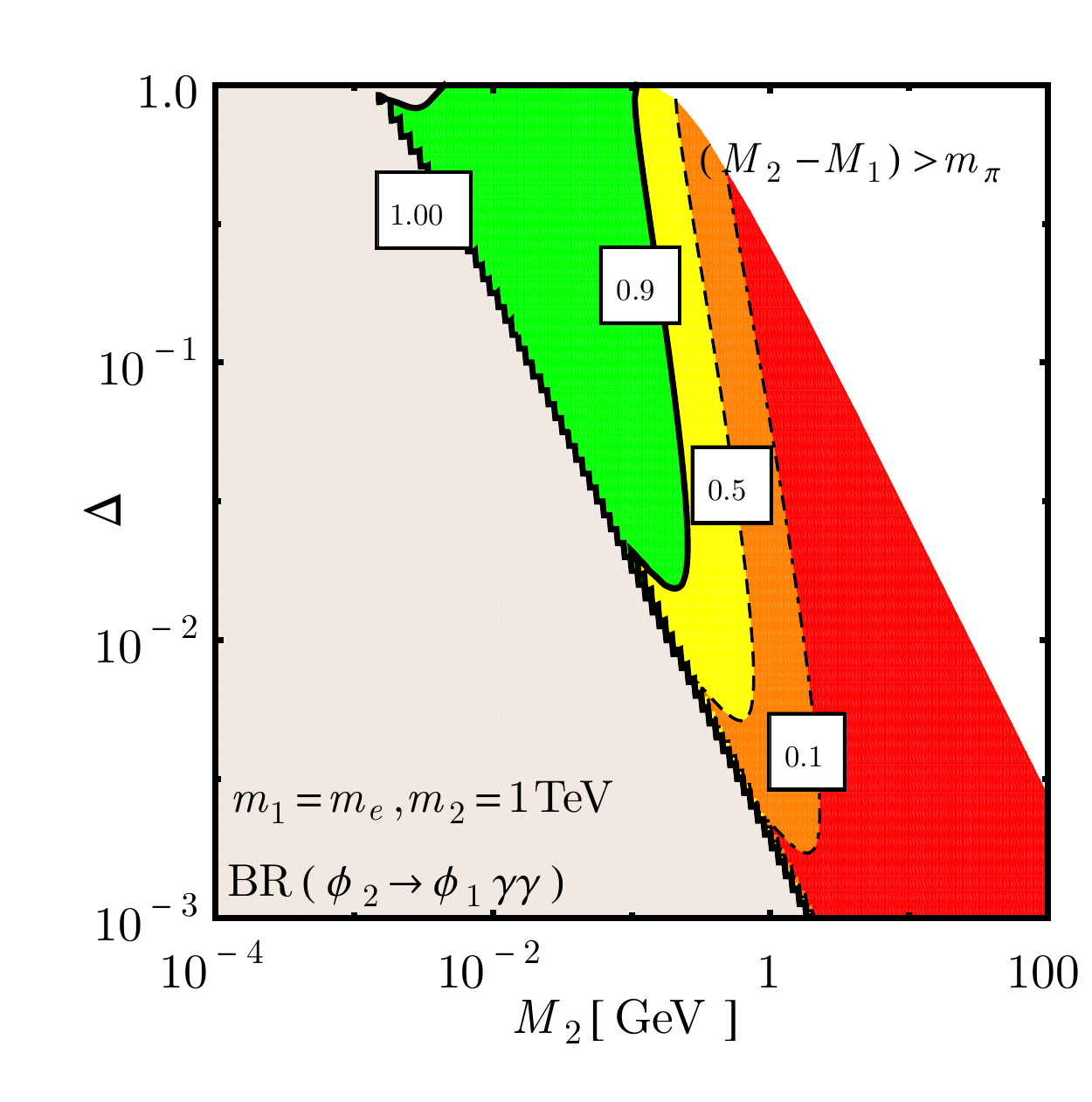} 
			\caption{Same as Fig.~\ref{fig:BRs}, but for the scenario described in Section \ref{sec:coupling-to-SM} For the plot we have assumed that the $Z_2$-even fermion is an electron and that the $Z_2$-odd fermion has mass $m_2=1$ TeV.}
		\label{fig:BR_lowm1}
	\end{center}
\end{figure}
	
The ratio of rates can be calculated from Eq.~\eqref{eq:approx_rates_SM4}. One finds that
\begin{eqnarray}
\frac{\Gamma_{\phi_1\gamma\gamma}}{\Gamma_{\phi_{1}f\bar f}}\,&\simeq&\,
\begin{cases}
\displaystyle{
\frac{\alpha^{2}_{\rm EM}\Delta^{2}}{2016\pi^{2}}\left(\frac{m_{2}}{m_{1}}
\right)^{2}\,\frac{\,{_{2}F}_1(3,4,8;\Delta)}{\,{_{2}F}_1(1,3,6;\Delta)}}
\hspace{5cm}{\rm for\,}M_{2}-M_{1}\,\ll\,2m_{1}\\
\!\begin{aligned}
&\displaystyle{
\frac{639\,\alpha^{2}_{\rm EM}}{20\pi^{2}\,\Delta^{2}}\left(\frac{m_1^2 m_2^2}{M^{4}_{2}}\right)}\times\\
&~~~~~\frac{\Big[\left(1+\frac{1}{200}\log^4\left(\frac{M^{2}_{2}\Delta^{2}}{2m^{2}_{1}}\right)\right){}_{2}F_{1}\left(1,2,4;\Delta\right)
	+0.03\,{\cal J}\left(M_{2},m_{1},\Delta\right)\Big]}{{}_{2}F_{1}\left(1,3,6;\Delta\,\right)}
\!\end{aligned}\\
\hspace{10.6cm}{\rm for\,}M_{2}-M_{1}\,\gg\,2m_{1}
\end{cases}\;.
\label{eqn:width_ratio_lowm1}
\end{eqnarray}
It follows from these expressions that $\phi_2\rightarrow\phi_1\gamma\gamma$ can have a sizable branching ratio, especially when $m_2\gg m_1, M_2$. This is illustrated in Fig.~\ref{fig:BR_lowm1}, 
which shows contour lines of the branching ratio for $\phi_2\rightarrow\phi_1\gamma\gamma$ for different values of $\Delta$ and $M_2$, for the specific case where the Standard Model fermion is an 
electron and the $Z_2$-odd exotic fermion has mass $m_2=1$ TeV.

\begin{figure}[t!]
	\begin{center}
		\includegraphics[width=0.49\textwidth]{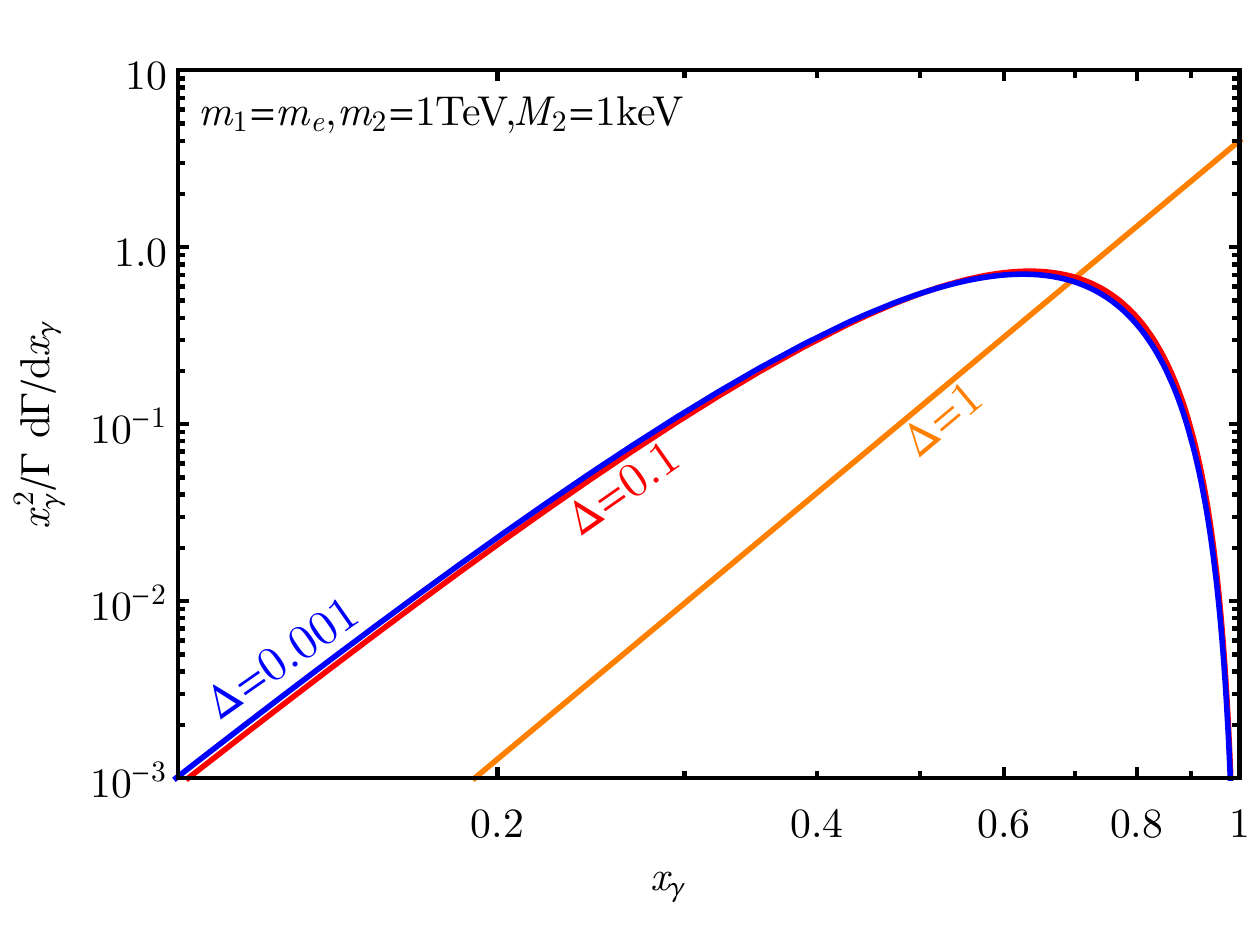}
		\includegraphics[width=0.49\textwidth]{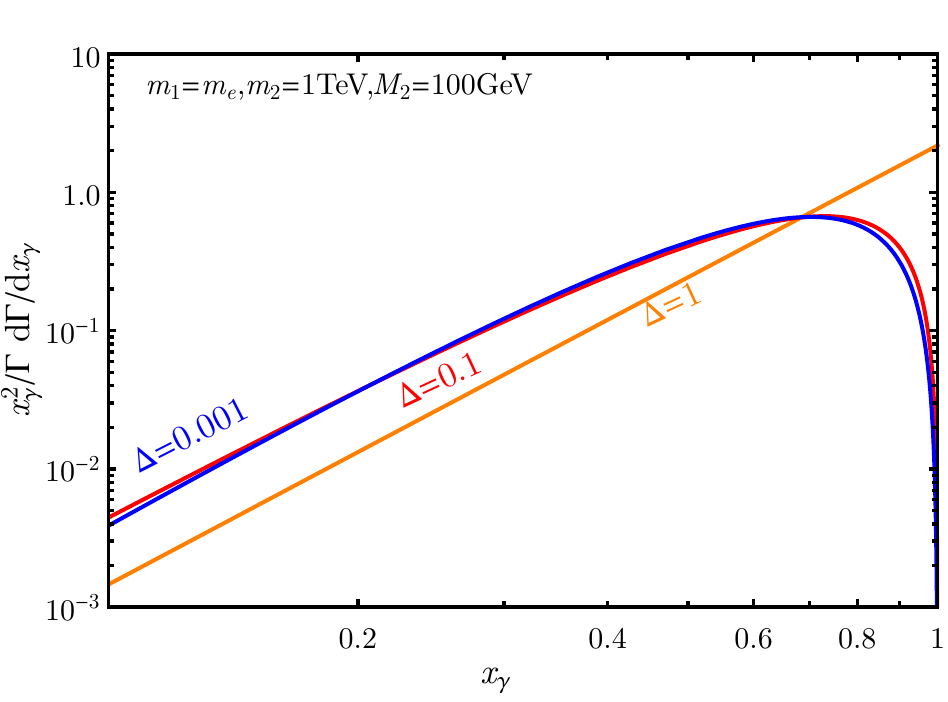}
		\caption{Same as Fig.~\ref{fig:diff_spectrum_eff}, but for the scenario described in Section \ref{sec:coupling-to-SM} For the plot we have assumed that the $Z_2$-even fermion is an electron, 
		that the $Z_2$-odd fermion has mass $m_2=1$ TeV, and that the decaying dark matter component mass is $M_2=1$ keV (left panel) and $M_2=100$ GeV (right panel).}
		\label{fig:diff_spectrum_SM}
	\end{center}
\end{figure}

\begin{figure}[t!]
	\begin{center}
		\includegraphics[width=0.49\textwidth]{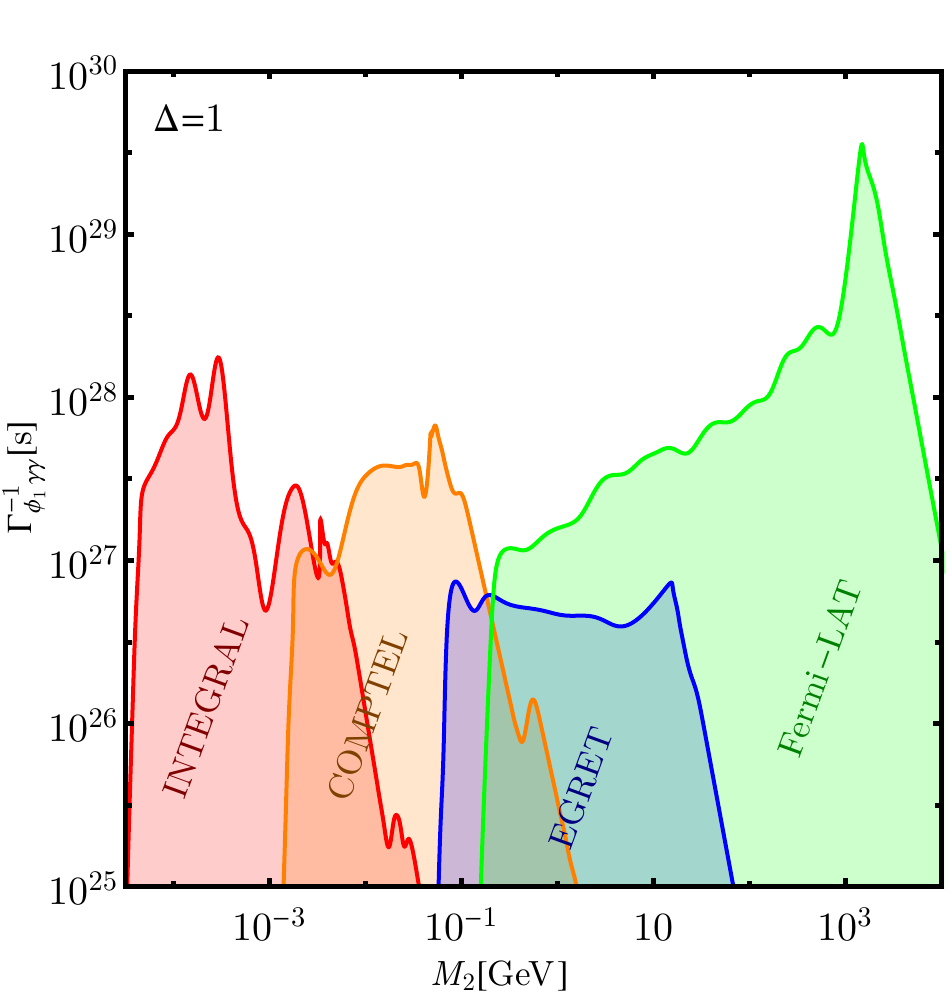}
		\includegraphics[width=0.49\textwidth]{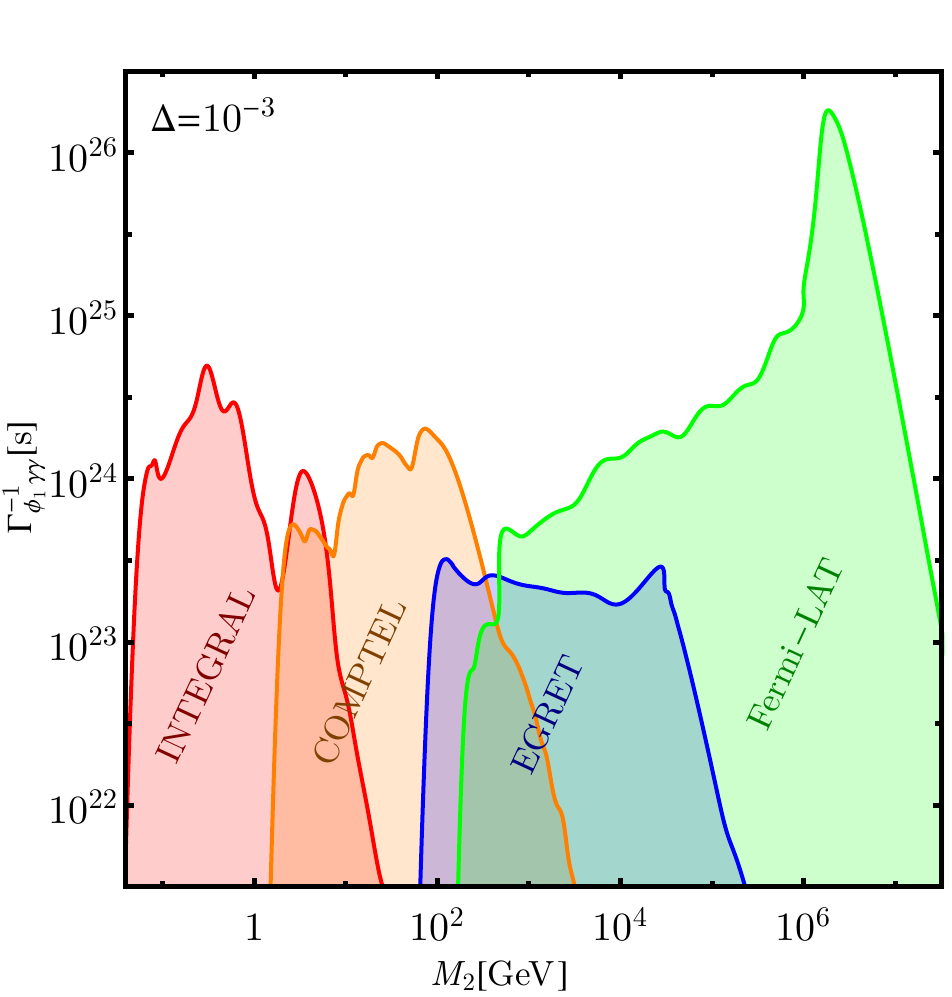}
		\caption{Lower limit on the inverse width of the decay process $\phi_{2}\rightarrow\phi_{1}\gamma\gamma$ as a function of the mass of the decaying dark matter component $M_2$, 
		for a very hierarchical ($\Delta=1$, left panel) and a very degenerate ($\Delta=10^{-3}$, right panel) dark matter mass spectrum with $\psi_1=$ electron for illustration. }
		\label{fig:lifetime_limit_loop}
	\end{center}
\end{figure}

The (normalized) differential photon spectrum is in this case
\begin{align}
\frac{1}{\Gamma_{\phi_1\gamma\gamma}}\frac{d\Gamma_{\phi_1\gamma\gamma}}{dx_{\gamma}}\,&\simeq&
\begin{cases}
\displaystyle{\frac{140}{{_{2}F}_1(3,4,8;\Delta)}\,\frac{x_{\gamma}^3(1-x_{\gamma})^{3}}{(1-x_{\gamma}\,\Delta)^{3}}} \hfill { \rm for\,}M_{2}-M_{1}\ll 2m_{1},
\\\displaystyle{\frac{x_{\gamma}(1-x_{\gamma})}{(1-\Delta\,x_{\gamma})}\,\frac{6\,+0.03\,\log^4\left[\left(\frac{2m^{2}_{1}}{M^{2}_{2}\Delta^2}\right)\frac{(1-\Delta\,x_{\gamma})}{x_{\gamma}(1-x_{\gamma})}\right]}{\Big[\left(1+\frac{1}{200}\log^4\left(\frac{M^{2}_{2}\Delta^{2}}{2m^{2}_{1}}\right)\right){}_{2}F_{1}\left(1,2,4;\Delta\right)+0.03{\cal J}\left(M_{2},m_{1},\Delta\right)\Big]}}\\
\hfill{\rm for\,}M_{2}-M_{1}\gg 2m_1
\end{cases}\;,
\label{eqn:2gamma-spectrum_lowm1_norm}
\end{align}
which is mostly dependent on the parameter $\Delta$, but also has a mild dependence on $M_2$. The differential photon spectrum is shown in Fig.~\ref{fig:diff_spectrum_SM}, 
taking for illustration $M_2=1$ keV and $M_2$=100 GeV. The differential spectrum is qualitatively similar to the one obtained in the effective theory approach analyzed in 
Section \ref{sec:eff_theory}, although shows some quantitative differences. The corresponding limits on the inverse width from gamma-ray telescopes are shown in Fig.~\ref{fig:lifetime_limit_loop}, 
for the representative cases $\Delta=1$ and $\Delta=10^{-3}$, for the case when the Standard Model fermion in the loop is an electron.

\section{A supersymmetric scenario: the right-chiral (s)neutrinos in $\snu\,$MSSM}
\label{sec:right_snu_DM}	

	The MSSM in its R-parity conserved form is a well-motivated 
	new physics scenario which not only solves the 
     {\it naturalness problem} but also provides a stable dark matter 
	candidate. A simple way of generating the 
	correct neutrino mass and mixing pattern is to extend the MSSM with
	right-handed neutrino superfields,  with tiny or vanishing Majorana masses, such that the total lepton number is approximately conserved and the neutrino masses are of Dirac-type.
	Along with the addition of right-handed neutrino superfields arises 
	the possibility of a new scalar dark sector comprising right 
	sneutrinos. Being SM gauge-singlets they can interact only through 
	their mixing with the left-handed partners through the neutrino Yukawa 
	coupling. Thus such sneutrinos are very weakly interacting~\cite{Asaka:2005cn,Asaka:2006fs,Asaka:2007zza,Ishiwata:2009gs,Banerjee:2016uyt,Gopalakrishna:2006kr,Banerjee:2018uut,Ghosh:2018hly,Choi:2018vdi}. 
	Neutrino oscillation data additionally requires
	the addition of at least two generations of right-handed neutrino 
	superfields. In case the sneutrino masses have a common origin at
	high-scale, they are likely to be nearly degenerate at the 
	electroweak scale, since their renormalization group equation
	is driven by the neutrino trilinear coupling~\cite{Martin:1997ns,Banerjee:2016uyt}
    which is usually proposed to be proportional to neutrino Yukawa 
    coupling $Y_{\nu}$, the proportionality constant being a SUSY
    breaking mass-scale $A_{\nu}$. 	
	
\begin{figure}[t!]
		\begin{center}
			\includegraphics[width=0.8\textwidth]{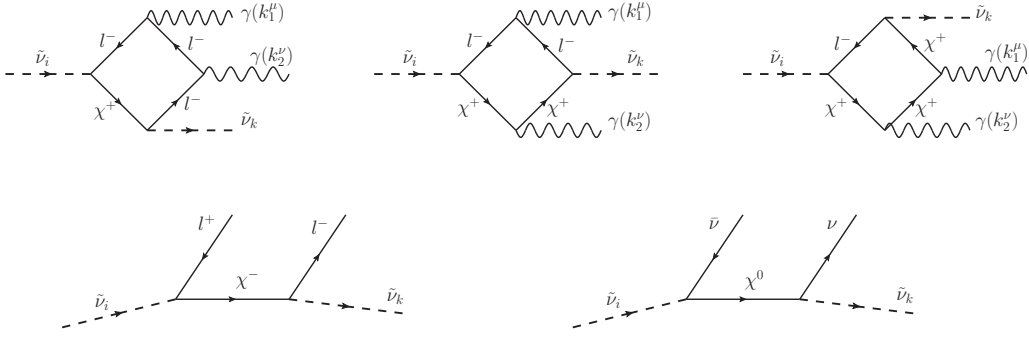}
			\vspace{1cm}
			\caption{Feynman diagrams giving the dominant contribution to the decay processes $\snu_{i}\rightarrow
			\snu_{k}\gamma\gamma$,$\snu_{i}\rightarrow\snu_{k}e^{-}e^{+}$ 
			and $\snu_{i}\rightarrow\snu_{k}\nu\bar{\nu}$. Diagrams where the photon lines are interchanged (not shown in the Figure) also contribute to the amplitudes.}
			\label{fig:right_snu_loops}
		\end{center}
	\end{figure}

 When the mass difference is smaller than twice the electron mass, the 
	heavier sneutrino ($\snu_{i}$) decays into the lighter sneutrino and 
	a neutrino-antineutrino pair via the exchange of virtual neutralinos, 
	or into the lighter sneutrino and two photons. For unitary sneutrino 
	mixing, it can be checked that the coupling $\snu_{i}\snu_{k}\,h\,(Z)$ is 
	forbidden for $i\neq k$ and hence the decays cannot occur via the 
	mediation of a virtual Higgs (Z) boson. The decay can however proceed at 
	the one loop level via diagrams such as the ones in 
	Fig.~\ref{fig:right_snu_loops}. Assuming that
	all supersymmetric particles, excepting the right-handed sneutrinos,
	are at the mass scale $m_{\rm SUSY}$, and  
	$m_{\snu_{R\,i}}\!\ll \!m_{\rm SUSY}$,  we checked that the dominant contribution to 
	the amplitude comes from lepton($l^{-}$)-chargino($\chi^{-}$) loops, 
	which is enhanced by a factor $\sim m_{\rm SUSY}/m_l$ compared to 
	other contributions. This scenario, therefore, falls into the class 
	of models analyzed in Section \ref{sec:coupling-to-SM}. Assuming a scheme
	of {\it minimal flavour violation}
	~\cite{Gabbiani:1996hi,DAmbrosio:2002vsn,Antonelli:2009ws,Buras:2010mh} in the leptonic sector, 
	the  Yukawa coupling of the sneutrino $i=1,2$ to the lepton $l=e,\mu,\tau$ is given by
	\begin{align}
	Y_{i l}= g_{W}\,{\rm sin}\,\Theta_{\snu_{ii}}\,U_{i\,l}\;,
	\label{eq:Yuks_SUSY}
	\end{align}
	where $g_{W}$ is weak gauge coupling, $U_{i\,l}$ are elements of the
	Pontecorvo-Maki-Nakagawa-Sakata (PMNS) matrix and $\Theta_{\snu_{ii}}$ is  the mixing angle between the left-sneutrino of flavor $i$ with the right-sneutrino of the same flavor, which reads
	\begin{equation}
	\tan\,\Theta_{\snu_{ii}}\,=\,\frac{2\,y_{\nu}v\,\sin\beta\,|\mu\cot\beta-A_{\nu}|}{m^{2}_{\snu_{L}}-m^{2}_{\snu_{R}}}\;.
	\label{eqn:snu_mixangle}
	\end{equation}
	Here, $y_{\nu}$ is neutrino
		Yukawa coupling which, for Dirac neutrinos, lies in the range $\simeq\,\left(2.8-4.4\right)
		\times\,10^{-13}$ where the lower and upper limit correspond respectively to a scenario of hierarchical and degenerate neutrino masses.
	
	Particularizing Eq.\eqref{eq:approx_rates_SM4} to  this model, and taking for simplicity $A_\nu, m_{\snu_{L}}\sim\,m_{\rm SUSY}\,\gg\,m_{\snu_{R}}$, with $m_{\rm SUSY}$ being the overall SUSY mass scale, 
	the partial rates in the different channels can be approximated as,
	\begin{align}
	\Gamma_{{\snu_{i}\rightarrow\snu_{k}\nu\bar \nu}}\,\simeq&\,\left(10^{26}{\rm s}\right)^{-1}\left(\dfrac{\underset{l,r=e,\mu,\tau}{\sum}{\rm Re}(Y_{il}Y^{*}_{kr})}{2.1\times10^{-17}}\right)^{2}\left(\frac{m_{\rm SUSY}}{1\,{\rm TeV}}\right)^{-4} \left(\dfrac{m_{\snu_{i}}}{1\,{\rm GeV}}\right)^{5}\Delta^{5},\nonumber\\
	\Gamma_{{\snu_{i}\rightarrow\snu_{k}e^{-}e^{+}}}\,\simeq&\,\left(10^{26}{\rm s}\right)^{-1}\left(\dfrac{{\rm Re}(Y_{ie}Y^{*}_{ke})}{2.1\times10^{-17}}\right)^{2}\left(\frac{m_{\rm SUSY}}{1\,{\rm TeV}}\right)^{-4} \left(\dfrac{m_{\snu_{i}}}{1\,{\rm GeV}}\right)^{5}\Delta^{5},\nonumber\\
	\Gamma_{\snu_{i}\rightarrow\snu_{k}\gamma\gamma} \simeq& \left(10^{26}{\rm s}\right)^{-1} \left(\frac{m_{\rm SUSY}}{1{\rm TeV}}\right)^{-2} \nonumber\\
&\times\begin{cases}	
\displaystyle{
\left(\dfrac{{\rm Re}(Y_{ie}Y^{*}_{ke})}{6.3\times10^{-20}}\right)^{2}\,\left(\dfrac{m_{\snu_{i}}}{1\,{\rm GeV}}\right)^{5}\Delta^{7}}
\hspace{4.8cm}\,{\rm for\,\,}m_{\snu_{i}}-m_{\snu_{k}}\!\ll\!2m_{e},\\
\!\begin{aligned}
\displaystyle{\left(\dfrac{{\rm Re}(Y_{ie}Y^*_{ke})}{6\times10^{-15}}\right)^{2}\,\left(\dfrac{m_{\snu_{i}}}{1\,{\rm GeV}}\right)\Delta^{3}\left[1+\frac{7}{5}\log^2\left(\frac{m^2_{\snu_i}\Delta^2}{2\,m^2_e}\right)\right]}
\!\end{aligned}
\hspace{1cm}\,{\rm for\,\,}m_{\snu_{i}}-m_{\snu_{k}}\!\gg\!2m_{e}\\
\end{cases}\;,
	\end{align}
	where we have assumed $\Delta \ll 1$.
	
When the mass difference between the sneutrinos is smaller than twice the electron mass, the only decays accessible are $\snu_{i}\rightarrow\snu_{k}\gamma\gamma$ and 
	$\snu_{i}\rightarrow\snu_{k}\nu\bar \nu$, with ratio of the rates approximately given by
	\begin{align}
	\frac{\Gamma_{\snu_{i}\rightarrow\snu_{k}\gamma\gamma}}{\Gamma_{\snu_{i}\rightarrow\snu_{k}\nu\bar \nu}}\simeq \frac{\alpha^{2}_{\rm EM}\Delta^{2}}{2016\pi^{2}}\left(\frac{U^{\dagger}_{ek}U_{ie}}{\underset{l,r=e,\mu,\tau}{\sum}U^{\dagger}_{rk}U_{il}}\right)\left(\frac{m^{2}_{\rm SUSY}}{m^{2}_{e}}\right)\;.
	\label{eq:diphoton_vs_invisible_SUSY}
	\end{align}
	
 In this regime one has $\Delta\leq 4 m_e/m_{\snu_{i}}$, therefore the diphoton decay channel can dominate over the ``invisible''  decay channel if the mass difference is 
	 not too small and if  $m_{\snu_{i}}\lessim 1.1\times\,10^{-4}\, m_{\rm SUSY}$ ; if the mass difference between the sneutrinos is generated through quantum effects by the 
	 tiny neutrino Yukawa coupling, such that $\Delta\ll 4 m_e/m_{\snu_{i}}$, then the decay will be dominated by the ``invisible'' channel.
	 
	When the right-sneutrino mass splitting is larger than twice the electron mass, $\Delta \geq 2 m_e/m_{\snu_{i}}$ the loop and $\alpha_{\rm EM}$ suppression factors in 
	Eq.~(\ref{eq:diphoton_vs_invisible_SUSY}) can be compensated by the (possibly large) factor $m_{\rm SUSY}^2/m^2_{\snu_{i}}$. In this regime, furthermore, the decay channel  
	$\snu_{i}\rightarrow\snu_{k}e^{-}e^{+}$ opens up. The ratio of the rates of $\snu_{i}\rightarrow\snu_{k}\gamma\gamma$ and $\snu_{i}\rightarrow\snu_{k}e^{+}e^{-}$ is given by
	\begin{eqnarray}
	\frac{\Gamma_{\snu_{i}\rightarrow\snu_{k}\gamma\gamma}}{\Gamma_{\snu_{i}\rightarrow\snu_{k}e^- e^+}}\,&\simeq&\,\frac{639\,\alpha^{2}_{\rm EM}}{40\pi^{2}}\left(\frac{m^{2}_{\rm SUSY}}{m^{2}_{\snu_{i}}}\right)\,{\cal F}\left(\frac{m^{2}_{\snu_{i}}\,\Delta^{2}}{2m^{2}_{e}}\right)\;,
	\label{eq:ratio_snus}
	\end{eqnarray}
	with ${\cal F}(x)=x^{-1} (1+7/5 \log^2x)$. This ratio is larger than 1 when the sneutrino is sufficiently light. On the other hand, in this regime $x\geq 2$, 
	which implies ${\cal F}(x)\,\lesssim\,0.93$. Therefore, the ratio Eq.~(\ref{eq:ratio_snus}) is necessarily smaller than 1 if $m_{\snu_i}\gtrsim 9\times\,10^{-3}\,m_{\rm SUSY}$.

For even larger masses, decays into a muon-antimuon 
pair, or mesons open up, usually taking a significant fraction of the total decay width. 
	
The Yukawa couplings inducing the sneutrino decay are, in simple scenarios, related to the sneutrino dark matter density. Concretely, for sneutrino dark matter generated by freeze-in of the 
slepton decays $\tilde{l}_{L}\rightarrow\,\snu_{R}W$ and $\snu_{L}\rightarrow\,\snu_{R}Z$, the relic abundance can be approximated as \cite{Asaka:2005cn},
\begin{equation}
	\Omega_{\snu_{i}}h^{2}\,\simeq\,0.12\,\left(\frac{g_{*}}{106.75}\right)^{-3/2}\,\left(\frac{{\rm sin}\Theta_{\snu_{ii}}}{6.08\times\,10^{-12}}\right)^{2}\left(\frac{{\rm sin}\beta}{1}\right)^{-2}\left(\frac{m_{\rm SUSY}}{1\,{\rm TeV}}\right)\left(\frac{m_{\snu_{i}}}{1\,{\rm GeV}}\right)\;,
\end{equation}
where $\Theta_{\snu_{ii}}$ was defined in Eq.~(\ref{eqn:snu_mixangle}), which in turn determines the Yukawa couplings of the model through Eq.~(\ref{eq:Yuks_SUSY}). 
We can then estimate the order of magnitude of the inverse width for the decay process $\snu_{i}\rightarrow\snu_{k}\gamma\gamma$ as 
\begin{align}
\Gamma_{\snu_{i}\rightarrow\snu_{k}\gamma\gamma} \simeq& \,\sin^4\beta \left(\frac{m_{\rm SUSY}}{1{\rm TeV}}\right)^{-4} \nonumber\\
&\times\begin{cases}	
	\displaystyle{
		\left(5\times 10^{33}{\rm s}\right)^{-1}\left(\dfrac{m_{\snu_{i}}}{1\,{\rm GeV}}\right)^{3}\Delta^{7}}&{\rm for\,\,}m_{\snu_{i}}-m_{\snu_{k}}\!\ll\!2m_{e},\\
		\displaystyle{\left(2\times 10^{36}{\rm s}\right)^{-1}\left(\dfrac{m_{\snu_{i}}}{1\,{\rm GeV}}\right)\Delta^{5}\,{\cal F}\left(\frac{m^2_{\snu_i}\Delta^2}{2\,m^2_e}\right)} & {\rm for\,\,}m_{\snu_{i}}-m_{\snu_{k}}\!\gg\!2m_{e}\\
\end{cases}\;,
\label{eq:snutime_relic}
\end{align}
where we have assumed that $\snu_i$ accounts for most of the dark matter of the Universe,  $|U_{ij}|\sim 1/\sqrt{3}$, $g_*=106.75$ and $\Delta\ll 1$. It is clear from eqn.~\eqref{eq:snutime_relic} that the 
choices of parameters favoured by freeze-in production  result in a diphoton 
flux which is well below the sensitivity of current or future gamma-ray telescopes. In variants of this scenario, however, the prospects of detection might be more promising.

\section{Summary and Conclusion}
\label{sec:conclusion}

We have considered a dark matter scenario where the lightest and next-to-lightest particles of the dark sector are singlet scalars, odd under a discrete $Z_2$ symmetry, 
while the Standard Model particles are all even. The lightest scalar, $\phi_1$, is assumed to be absolutely stable. However, the next-to-lightest scalar, $\phi_2$, could decay 
into the lightest one together with Standard Model particles. 

We have investigated the gamma-ray signatures produced in the decay. Angular momentum conservation forbids the decay 
$\phi_2\rightarrow \phi_1 \gamma$, hence we have focused on the three-body decay  $\phi_2\rightarrow \phi_1 \gamma\gamma$, which presents a characteristic energy spectrum with a peak 
close to the kinematic end-point and which could be easily distinguished from the (featureless) astrophysical diffuse gamma-ray background. 

We have constructed the most general CP-conserving effective Lagrangian inducing this decay either via a Higgs portal interaction or via dimension six operators, 
and we have proposed a UV complete model that generates those dimension six operators involving one $Z_2$-even and one $Z_2$-odd heavy fermion running in the loops. 
We have calculated the decay rates for the processes $\phi_2\rightarrow \phi_1 \gamma\gamma$, $\phi_2\rightarrow \phi_1 e^- e^+$ and $\phi_2\rightarrow \phi_1 \nu \bar \nu$ and  
identified the regions of the parameter space where the branching ratio of $\phi_2\rightarrow \phi_1 \gamma\gamma$ is sizable. 
We have also derived stringent limits on the inverse width for this process from the non-observation of such gamma-ray feature in the diffuse gamma-ray background inferred 
from the INTEGRAL, COMPTEL, EGRET and Fermi-LAT observations, both for large and for small mass differences between the two dark matter components. These limits in turn translate into stringent limits on the parameters of the model.

We have also analyzed a variant of our UV complete model where the $Z_2$-even fermion is identified with a Standard Model fermion. 
For this scenario, the effective theory approach to the decay $\phi_2\rightarrow \phi_1 \gamma\gamma$ is not valid and therefore requires a separate analysis. 
In particular, we show that the photon energy spectrum (and accordingly the limits on the inverse width from gamma-ray data) differs from the one calculated in the effective theory approach. 
As a particular case of this scenario, we have considered the MSSM augmented by three right-handed neutrino superfields, 
and we have studied the gamma-ray signals generated in the decay of the next-to-lightest supersymmetric particle into the lightest in the case where the total lepton number is conserved. 
For freeze-in production the predicted gamma-ray flux is unfortunately many orders of magnitude below the sensitivity of current or future gamma-ray telescopes, due to the smallness of the 
neutrino Yukawa coupling. In other scenarios, however, the gamma-ray signal from $\phi_2\rightarrow \phi_1\gamma\gamma$ may be within the reach of experiments.

\section*{Acknowledgements}
\label{sec:Acknowledgements}
The work of AG, TM and BM was partially supported by funding 
available from the Department of Atomic Energy, Government of India, 
Regional Centre for Accelerator-based Particle Physics (RECAPP), 
Harish-Chandra Research Institute. 
The work of AI was partially supported by the DFG cluster of excellence ORIGINS and by the Collaborative Research Center SFB1258. 
The work of BM was partially supported by the DFG cluster of excellence `Origin and Structure of the Universe'.

\bibliographystyle{JHEP-mod}
\bibliography{references}
\end{document}